\documentclass[lettersize,journal]{IEEEtran}
\usepackage{amsmath,amssymb,amsfonts}
\usepackage{array}
\usepackage[caption=false,font=normalsize,labelfont=sf,textfont=sf]{subfig}
\usepackage{textcomp}
\usepackage{stfloats}
\usepackage{url}
\usepackage{verbatim}
\usepackage{graphicx}
\usepackage{cite}
\usepackage{multirow}
\usepackage{booktabs}
\usepackage{xcolor}
\usepackage{makecell}
\usepackage{booktabs,dcolumn}
\usepackage{nicematrix}

    \def\Complex{{\rm\rule[.23ex]{.03em}{1.1ex}\kern-.3em{C}}}

    \newcommand{\be}{\begin{equation}} \newcommand{\ee}{\end{equation}}
    \newcommand{\bea}{\begin{eqnarray}} \newcommand{\eea}{\end{eqnarray}}
    \newcommand{\benum}{\begin{enumerate}} \newcommand{\eenum}{\end{enumerate}}

    \newtheorem{Theorem}{Theorem}

    \newtheorem{Lemma}{Lemma}

    \newcommand{\qa}{{\bf a}}

    \newcommand{\qh}{{\bf h}}

    \newcommand{\qu}{{\bf u}}
    \newcommand{\qv}{{\bf v}}
    
    \newcommand{\qx}{{\bf x}}
    \newcommand{\qy}{{\bf y}}
    \newcommand{\qz}{{\bf z}}

    \newcommand{\qA}{{\bf A}}
    \newcommand{\qB}{{\bf B}}
    
    \newcommand{\qD}{{\bf D}}

    \newcommand{\qH}{{\bf H}}
    \newcommand{\qI}{{\bf I}}

    \newcommand{\qQ}{{\bf Q}}
    \newcommand{\qR}{{\bf R}}

    \newcommand{\qU}{{\bf U}}
    \newcommand{\qV}{{\bf V}}

    \newcommand{\qzero}{{\bf 0}}
    \newcommand{\qone}{{\bf 1}}

    \newcommand{\qPhi}{{\boldsymbol \Phi}}

    \newcommand{\qSigma}{{\boldsymbol \Sigma}}

    \newcommand{\qOmega}{{\boldsymbol \Omega}}

    \newcommand{\qomega}{{\boldsymbol \omega}}
    \newcommand{\qphi}{{\boldsymbol\phi}}

    \newcommand{\bbR}{{\mathbb R}}
    \newcommand{\bbC}{{\mathbb C}}

    \newcommand{\diag}{{\sf diag}}
    
    \newcommand{\tr}{{\sf tr}}
    
    \newcommand{\rank}{{\sf rank}}

    \newcommand{\rl}[1]{\color{red}#1}

\begin{document}

\title{End-User-Centric Collaborative MIMO: Performance Analysis and Proof of Concept}

\author{
Chao-Kai~Wen,~\IEEEmembership{Fellow,~IEEE},~Yen-Cheng~Chan,~Tzu-Hao~Huang,
~Hao-Jun~Zeng,~Fu-Kang~Wang,~\IEEEmembership{Member,~IEEE},
Lung-Sheng~Tsai,~and~Pei-Kai~Liao,~\IEEEmembership{Senior Member,~IEEE}

\thanks{{C.-K.~Wen}, Y.-C. Chan, and T.-H. Huang are with the Institute of Communications Engineering, National Sun Yat-sen University, Kaohsiung 80424, Taiwan, Email: chaokai.wen@mail.nsysu.edu.tw, \{ycchan7250, peter94135\}@gmail.com.}

\thanks{C.-H.~Wu and F.-K.~Wang are with the Department of Electrical Engineering, National Sun Yat-sen University, Kaohsiung 804, Taiwan, Email: sifwadov5656@gmail.com, fkw@mail.ee.nsysu.edu.tw.}

\thanks{{L.-S.~Tsai} and {P.-K.~Liao} are with the MediaTek Inc., Hsinchu, Taiwan, Email: \{Longson.Tsai, pk.liao\}@mediatek.com.}

}

\maketitle

\begin{abstract}
The trend toward using increasingly large arrays of antenna elements continues. However, fitting more antennas into the limited space available on user equipment (UE) within the currently popular Frequency Range 1 spectrum presents a significant challenge. This limitation constrains the capacity-scaling gains for end users, even when networks support a higher number of antennas. To address this issue, we explore a user-centric collaborative MIMO approach, termed UE-CoMIMO, which leverages several fixed or portable devices within a personal area to form a virtually expanded antenna array. This paper develops a comprehensive mathematical framework to analyze the performance of UE-CoMIMO. Our analytical results demonstrate that UE-CoMIMO can significantly enhance the system's effective channel response within the current communication system without requiring extensive modifications. Further performance improvements can be achieved by optimizing the phase shifters on the expanded antenna arrays at the collaborative devices. These findings are corroborated by ray-tracing simulations. Beyond the simulations, we implemented these collaborative devices and successfully conducted over-the-air validation in a real 5G environment, showcasing the practical potential of UE-CoMIMO. Several practical perspectives are discussed, highlighting the feasibility and benefits of this approach in real-world scenarios.
\end{abstract}

\begin{IEEEkeywords}
Device Collaborative MIMO, Spectral Efficiency, Proof of Concept
\end{IEEEkeywords}

\section{Introduction}

\IEEEPARstart{M}{assive} MIMO is a key technology for enhancing spectral efficiency in wireless networks. Full Dimension MIMO (FD-MIMO), which utilizes large antenna arrays, was first introduced in 3GPP Release 13 for LTE-Advanced Pro (4G). This foundation paved the way for the standardization of massive MIMO in 5G New Radio (NR) with 3GPP Release 15. Subsequent releases have further refined this technology, significantly increasing the number of ports for channel state information (CSI) reporting, from 32 to 128 in Release 19 \cite{Chen-23JSAC,Lin-23ArXiv}. Another approach to achieving massive MIMO is through the aggregation of antennas from multiple transmission/reception points (multi-TRPs) \cite{Jin-23JSAC}. This method enhances network capacity and reliability by leveraging spatial diversity from multiple locations. Looking ahead to 6G, the trend of increasing antenna elements on both the centralized and distributed network sides is expected to continue \cite{Wen-24COMSTMAG,Bjornson-24ArXiv}.

Efforts to maximize antenna numbers on the user equipment (UE) side have also been pursued \cite{Wong-23ACCESS}. However, physical limitations restrict the number of antennas that UEs, such as smartphones and wearable devices, can accommodate, typically limiting them to 4 to 8 antennas within the Frequency Range 1 (FR1) spectrum. This constraint hinders MIMO capacity scaling and reduces the number of spatial streams that a UE can support, making carrier aggregation the primary method for addressing throughput demands under these limitations.

\begin{figure}
\centering
\includegraphics[width=3.50in]{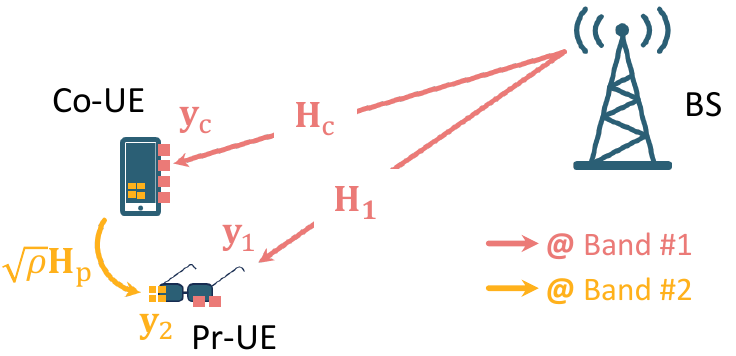}
\caption{A UE-CoMIMO System.}
\label{fig:UE-CoMIMO}
\vspace{-0.3cm}
\end{figure}

In recent years, the proliferation of personal devices such as smartphones, wearable devices, and portable chargers within personal area networks has created opportunities for deeper physical layer collaboration, effectively increasing the number of available antenna elements. This concept, known as end-user-centric collaborative MIMO (UE-CoMIMO) \cite{Tsai-23COMMAG}, was proposed in 3GPP \cite{RWS230111}. It enables multiple devices to collaboratively process their received signals, thereby augmenting the number of antennas and enhancing the number of spatial streams.

As illustrated in Fig.~\ref{fig:UE-CoMIMO}, in a UE-CoMIMO setup, the base station (BS) simultaneously transmits data to both the primary UE and a nearby collaborative UE using a low-frequency band. The collaborative UE then amplifies and forwards its received signal to the primary UE via a high-frequency band. The primary UE can either utilize additional high-frequency antennas or share antennas across both low- and high-frequency bands. This flexibility is further enhanced by the smaller size of high-frequency antennas, making them easier to integrate into the compact spaces of terminal devices. This configuration effectively increases the number of antennas available to the primary UE. For example, extended reality glasses equipped with two antennas could collaborate with a smartphone that has four antennas, resulting in a combined six-antenna system capable of supporting higher-order MIMO ranks.

While the concept of frequency-translation amplify-and-forward (AF) relays has previously been proposed to assist antenna-limited UEs \cite{Ng-16TWC, Sanchez-19TVT}, prior approaches have primarily utilized frequency division multiplexing to enhance spatial diversity rather than focusing on spatial multiplexing. Additionally, although UE-CoMIMO \cite{Tsai-23COMMAG} has demonstrated system-level gains in both spatial diversity and multiplexing, it lacks a comprehensive mathematical formulation and dedicated optimization algorithms. To address this gap, \cite{Chen-24} recently formulated a rate maximization problem for the UE-CoMIMO system and proposed optimization algorithms for relay amplifying matrices. However, the study assumes fully connected relay structures between low- and high-frequency antennas, which allow both amplitude and phase adjustments. Such assumptions may be impractical for terminal devices due to hardware constraints. Additionally, the proposed approach relies on full knowledge of individual channel states for the first and second relay hops, significantly increasing the complexity and computational burden on the collaborative UE. These challenges raise doubts about the practical feasibility of the UE-CoMIMO system.

In this paper, we address these challenges and aim to advance the practical development of the UE-CoMIMO system for real-world applications in personal area networks. Specifically, we have implemented the UE-CoMIMO system and conducted real-world testing. Our contributions are summarized as follows:
\begin{itemize}
  \item {\bf Theoretical Performance Analysis of the UE-CoMIMO System:}
  We provide a theoretical framework for analyzing the spectral efficiency (SE) of the UE-CoMIMO system, focusing on the singular values of its reformulated channel matrix. This matrix is constructed by incorporating the relay link's channel matrix as an additional row to the original channel matrix from the BS. Our analysis shows that incorporating the relay link consistently enhances the singular values, thereby improving system performance even without phase optimization. This is in contrast to reconfigurable intelligent surfaces (RIS), where performance gains rely heavily on precise phase optimization \cite{Xie-23TCOM,Shaikh-23SJ,Chen-23TVT,Tang-20JSAC}.

 \item {\bf Optimization Algorithms for Relay Phase Adjustment:}
 Prior work \cite{Sanchez-19TVT,Chen-24} assumes fully connected relay structures between low- and high-frequency antennas, allowing both amplitude and phase adjustments, which may be impractical for terminal devices. In this paper, we explore simpler relay structures, such as parallel phase-adjustable links between low- and high-frequency antennas (multiplexing structure) or a single link through a phase combiner/splitter between low- and high-frequency antennas (diversity structure). We develop optimization algorithms for phase adjustment in these structures that efficiently achieve near-optimal settings without requiring full CSI of the relay link. This approach results in a straightforward protocol and a practical, implementable algorithm.

 \item {\bf Simulations and Experimental Validation:}
 The performance of the UE-CoMIMO system is sensitive to the relative positioning of collaborating devices. We conducted ray-tracing simulations in both indoor and outdoor environments, revealing that different relay structures are optimal for different use cases. These simulations provided valuable insights into system design. Additionally, we implemented the UE-CoMIMO system in real-world settings to validate the simulation results and demonstrate its practical feasibility.
\end{itemize}

The remainder of this paper is organized as follows: Section \ref{sec:System_Model} presents the system model for the UE-CoMIMO system. Section \ref{sec:Perform_Analysis} analyzes the SE of the system and develops algorithms for optimizing the relay phases. Section \ref{sec:Sim&Dis} includes extensive simulations, such as ray-tracing, to explore the properties of optimal phases and their performance in various scenarios. Section \ref{sec:Experiments} discusses the system implementation and evaluates its performance through over-the-air testing. Finally, Section \ref{sec:Con} concludes the paper.


\section{System Model}
\label{sec:System_Model}

As shown in Fig.~\ref{fig:UE-CoMIMO}, we consider a MIMO system comprising a BS and a primary UE (Pr-UE). The transmission from the BS to Pr-UE occurs through the $f_{\rm L}$ frequency band. The transmitted signal, denoted as $\qx$, propagates through the wireless channel and is received by both the Pr-UE and a collaborative UE (Co-UE). The received signals at the Pr-UE and Co-UE are expressed as
\begin{subequations}
\begin{align}
 \qy_1 &= \qH_1 \qx + \qz_1, \\
 \qy_{\rm c} &= \qH_{\rm c} \qx + \qz_{\rm c}.
\end{align}
\end{subequations}
Here, $\qH_1 \in \bbC^{N_1 \times M}$ represents the channel matrix between the BS and the Pr-UE (direct link) with $M$ transmit antennas and $N_1$ receive antennas. Similarly, $\qH_{\rm c} \in \bbC^{N_{\rm c} \times M}$ denotes the channel matrix between the BS and Co-UE (first-hop channel), where $N_{\rm c}$ is the number of receive antennas at the Co-UE. The terms $\qz_1 \in \bbC^{N_1 \times 1}$ and $\qz_{\rm c} \in \bbC^{N_{\rm c} \times 1}$ are complex Gaussian noise vectors at the Pr-UE and Co-UE, respectively.

The Co-UE employs a frequency-translation AF repeater to relay its received signal, $\qy_{\rm c}$, to the Pr-UE via another frequency band, $f_{\rm H}$. At the Pr-UE, signals from the $f_{\rm H}$ band are received using $N_2$ antennas. The signal forwarded by the Co-UE contributes to the received signal at the Pr-UE, given as
\begin{equation} \label{eq:y_2}
 \qy_2 = \sqrt{\rho} \, \qH_{\rm p}  \qy_{\rm c} + \qz_2 \approx \sqrt{\rho} \, \qH_{\rm p} \qH_{\rm c} \qx  + \qz_2,
\end{equation}
where $\qH_{\rm p} \in \bbC^{N_2 \times N_{\rm c}}$ denotes the relay channel between the Co-UE and the Pr-UE, and $\rho$ represents the power gain from the low-noise amplifier (LNA) for the second-hop link. The term $\qz_2 \in \bbC^{N_2 \times 1}$ represents complex Gaussian noise at the Pr-UE. Unlike a conventional repeater with a high amplification gain (e.g., 90-100 dB), the Co-UE uses LNAs with low amplification (a few dB). Thus, the noise contribution of $\qz_{\rm c}$ is negligible and can be ignored, resulting in the approximation in \eqref{eq:y_2}.

For simplicity, we define $\qH_2 = \sqrt{\rho}\, \qH_{\rm p} \qH_{\rm c} \in \bbC^{N_2 \times M}$ and consolidate the signals and noise as follows:
\begin{equation} \label{eq:def_H,y,z}
 \qH = \begin{bmatrix} \qH_1 \\ \qH_2 \end{bmatrix}, ~\qy = \begin{bmatrix} \qy_1 \\ \qy_2 \end{bmatrix}, ~\qz = \begin{bmatrix} \qz_1 \\ \qz_2 \end{bmatrix}.
\end{equation}
Here, $\qH \in \bbC^{(N_1+N_2) \times M}$ and $\qy, \qz \in \bbC^{(N_1+N_2) \times 1}$. The Pr-UE can receive signals from both the $f_{\rm L}$ and $f_{\rm H}$ bands simultaneously.
As a result, the overall input-output relationship of the system can now be compactly expressed as
\begin{equation}
 \qy = \qH \qx +  \qz.
\end{equation}
For analysis and without loss of generality, we assume that the noise vector $\qz$ has been pre-whitened, so it represents a standard complex Gaussian noise vector.

It is important to note that UE-CoMIMO differs from direct communication using dual spectrum. The BS does not transmit simultaneously on both $f_{\rm L}$ and $f_{\rm H}$ bands in UE-CoMIMO \cite{Tsai-23COMMAG}. Instead, the Co-UE, typically located close to the Pr-UE and operating with low transmission power, minimizes network interference. From the network perspective, the BS communicates with an aggregated user (formed by multiple devices) on the $f_{\rm L}$ band while the $f_{\rm H}$ band can still serve other UEs.


\section{Performance Analysis and Customize $\qH_2$}
\label{sec:Perform_Analysis}

The SE of the UE-CoMIMO system highly depends on the singular values of $\qH$, which are formed by appending $\qH_2$ as a row to $\qH_1$.
We express the singular value decomposition (SVD) of $\qH_1$ as
\begin{equation}
  \qH_1 = \qU_{{\rm H}_1} \qSigma_{{\rm H}_1} \qV_{{\rm H}_1}^{\rm H},
\end{equation}
where $\qU_{{\rm H}_1} \in \bbC^{N_1 \times N_1}$ and $\qV_{{\rm H}_1} \in \bbC^{M \times M}$ are unitary matrices, $\qSigma_{{\rm H}_1} \in \bbR^{N_1 \times M}$ is zero except on the main diagonal, which contains non-negative entries in decreasing order. The columns of $\qU_{{\rm H}_1}$ and $\qV_{{\rm H}_1}$ are the left singular vectors and the right singular vectors of $\qH_1$, respectively; the diagonal entries of $\qSigma_{{\rm H}_1}$ are the singular values of $\qH_1$, arranged as follows:
\begin{equation}
  \sigma_{1}(\qH_1)  \geq \cdots \geq \sigma_{M-1}(\qH_1) \geq  \sigma_{M}(\qH_1)  \geq 0.
\end{equation}
Similarly, we can express the SVD of $\qH$ as
\begin{equation}
  \qH = \qU_{{\rm H}} \qSigma_{{\rm H}} \qV_{{\rm H}}^{\rm H},
\end{equation}
where $\qU_{{\rm H}} \in \bbC^{(N_1+N_2) \times (N_1+N_2)}$, $\qV_{{\rm H}} \in \bbC^{M \times M}$ are unitary matrices, and $\qSigma_{{\rm H}} \in \bbR^{(N_1+N_2) \times M}$ with the real numbers
\begin{equation}
   \sigma_{1}(\qH) \geq \cdots \geq \sigma_{M-1}(\qH) \geq  \sigma_{M}(\qH)  \geq 0
\end{equation}
on the diagonal.

This section explains how the singular values of $\qH$ can be modified from those of $\qH_1$ using $\qH_2$. We begin by demonstrating that the singular values of $\qH$ can be represented as a rank-$N_2$ modification of the singular values of $\qH_1$ in the following subsection. Then, we explicitly express the singular value update in terms of $\qH_1$ and $\qH_2$. Finally, we illustrate how to customize the channel $\qH_2$ by introducing phase shifters (PSs) and develop an efficient algorithm for this purpose.
 
\subsection{Rank-$N_2$ Modification}\label{sec:Rank Modification}

First, consider the case $N_1 < M$. We define the following matrices:
\begin{equation} \label{eq:def_V_H1in12}
 \qV_{{{\rm H}_1}}  = \begin{bmatrix} \qV_{{{\rm H}_1},1}  ~\qV_{{{\rm H}_1},2} \end{bmatrix}~\mbox{and}~
 \qSigma_{{\rm H}_1} = \begin{bmatrix} \qD_{{\rm H}_1}  ~\qzero  \end{bmatrix},
\end{equation}
where $\qV_{{{\rm H}_1},1} \in \bbC^{M \times N_1}$, $\qV_{{{\rm H}_1},2} \in \bbC^{M \times (M-N_1)}$, and $\qD_{{\rm H}_1} \in \bbR^{N_1 \times N_1}$. Here, $\qD_{{\rm H}_1}$ is a diagonal matrix containing the nonzero singular values of $\qH_1$. We further define $\qH'_{2,1} = \qH_2 \qV_{{{\rm H}_1},1} \in \bbC^{N_2 \times N_1}$ and $\qH'_{2,2} = \qH_2 \qV_{{{\rm H}_1},2} \in \bbC^{N_2 \times (M-N_1)}$. By substituting \eqref{eq:def_V_H1in12} into the definition of $\qH$ in \eqref{eq:def_H,y,z} and using the above definitions, $\qH$ can be expressed as:
\begin{align}
 \qH &= \begin{bmatrix}
        \qU_{{\rm H}_1} & \qzero  \\
        \qzero   & \qI_{N_2}
       \end{bmatrix}
       \begin{bmatrix}
       \qD_{{\rm H}_1}  & \qzero \\
       \qH'_{2,1} &  \qH'_{2,2}
       \end{bmatrix}
       \begin{bmatrix}
       \qV_{{{\rm H}_1},1}^{\rm H} \\
       \qV_{{{\rm H}_1},2}^{\rm H}
       \end{bmatrix} \notag \\
    &= \begin{bmatrix}
        \qU_{{\rm H}_1} & \qzero \\
        \qzero   & \qI_{N_2}
       \end{bmatrix}
       \begin{bmatrix}
       \qD_{{\rm H}_1}  & \qzero & \qzero \\
       \qH'_{2,1} &  \qR'_{2,2} & \qzero
       \end{bmatrix}
       \begin{bmatrix}
       \qV_{{{\rm H}_1},1}^{\rm H} \\
       (\qV_{{{\rm H}_1},2} \qQ'_{2,2} )^{\rm H}
       \end{bmatrix}, \label{eq:H_inOtherForm_N<M}
\end{align}
where $\qH'_{2,2} \qQ'_{2,2} = \qR'_{2,2}$ is the QR decomposition. Here, $\qQ'_{2,2} \in \bbC^{(M-N_1) \times N_2} $ is a matrix with orthogonal columns, and $\qR'_{2,2} \in \bbC^{N_2 \times N_2}$ is a lower triangular matrix. Define
\begin{equation} \label{eq:A_def_N<M}
 \qA = \begin{bmatrix}
       \qD_{1} & \qzero   \\
       \qH'_{2,1} &  \qR'_{2,2}
       \end{bmatrix}
       \in \bbC^{(N_1+N_2) \times (N_1+N_2)}.
\end{equation}
Since both sides of \eqref{eq:H_inOtherForm_N<M} are unitary matrices, finding the singular values of $\qH$ reduces to computing the SVD of $\qA$.

Specifically, the SVD of $\qA$ is given by
\begin{equation} \label{eq:A_svd_N<M}
 \qA = \qU_{{\rm A}}
       \begin{bmatrix}
       \qD_{\rm A} & \qzero
       \end{bmatrix}
       \begin{bmatrix}
       \qV_{{\rm A},1}^{\rm H} \\
       \qV_{{\rm A},2}^{\rm H}
       \end{bmatrix}  ,
\end{equation}
where $\qU_{{\rm A}} \in \bbC^{(N_1+N_2) \times (N_1+N_2)}$, $\qD_{\rm A} \in \bbR^{(N_1+N_2) \times (N_1+N_2)}$, $\qV_{{\rm A},1} \in \bbC^{M \times (N_1+N_2)}$, and $\qV_{{\rm A},2} \in \bbC^{M \times M-(N_1+N_2)}$.\footnote{The SVD of $\qH$ is
\begin{equation*}
 \qH = \qU_{{\rm H}}
       \begin{bmatrix}
       \qD_{\rm H} & \qzero
       \end{bmatrix}
       \begin{bmatrix}
       \qV_{{\rm H},1}^{\rm H} \\
       \qV_{{\rm H},2}^{\rm H}
       \end{bmatrix} ,
\end{equation*}
where
\begin{align*}
 \qU_{{\rm H}} &=
        \begin{bmatrix}
        \qU_{{\rm H}_1} & \qzero  \\
        \qzero   & \qI_{N_2}
       \end{bmatrix}
        \qU_{{\rm A}}, \\
 \qV_{{\rm H},1} &=  \begin{bmatrix} \qV_{{{\rm H}_1},1} & \qV' \end{bmatrix} \qV_{{\rm A},1}.
\end{align*}
Here, $\qV'$ and $\qV_{{\rm H},2}$ represent the first $N_2$ columns and remaining columns of $\qV_{{{\rm H}_1},2} \qQ'_{2,2}$, respectively.}
Let $\qH'_2 = [\qH'_{2,1}  ~ \qR'_{2,2}]$. By combining \eqref{eq:A_def_N<M} and \eqref{eq:A_svd_N<M}, we obtain
\begin{equation} \label{eq:AA_InUpdate_2}
 \qA^{\rm H} \qA =
       \begin{bmatrix}
        \qD_{{\rm H}_1}^2 & \qzero  \\
        \qzero   & \qzero
       \end{bmatrix} + (\qH'_2)^{ {\rm H}} \qH'_2 = \qV_{{\rm A}} \qD_{\rm A}^2 \qV_{{\rm A}}^{\rm H}.
\end{equation}
This represents a rank-$N_2$ modification of $\qD_{{\rm H}_1}^2$.

A similar approach applies to the case $N_1 \geq M$. For further details, refer to Appendix A.

\subsection{Singular Value Update}

\begin{figure*}
\centering
\includegraphics[width=7.00in]{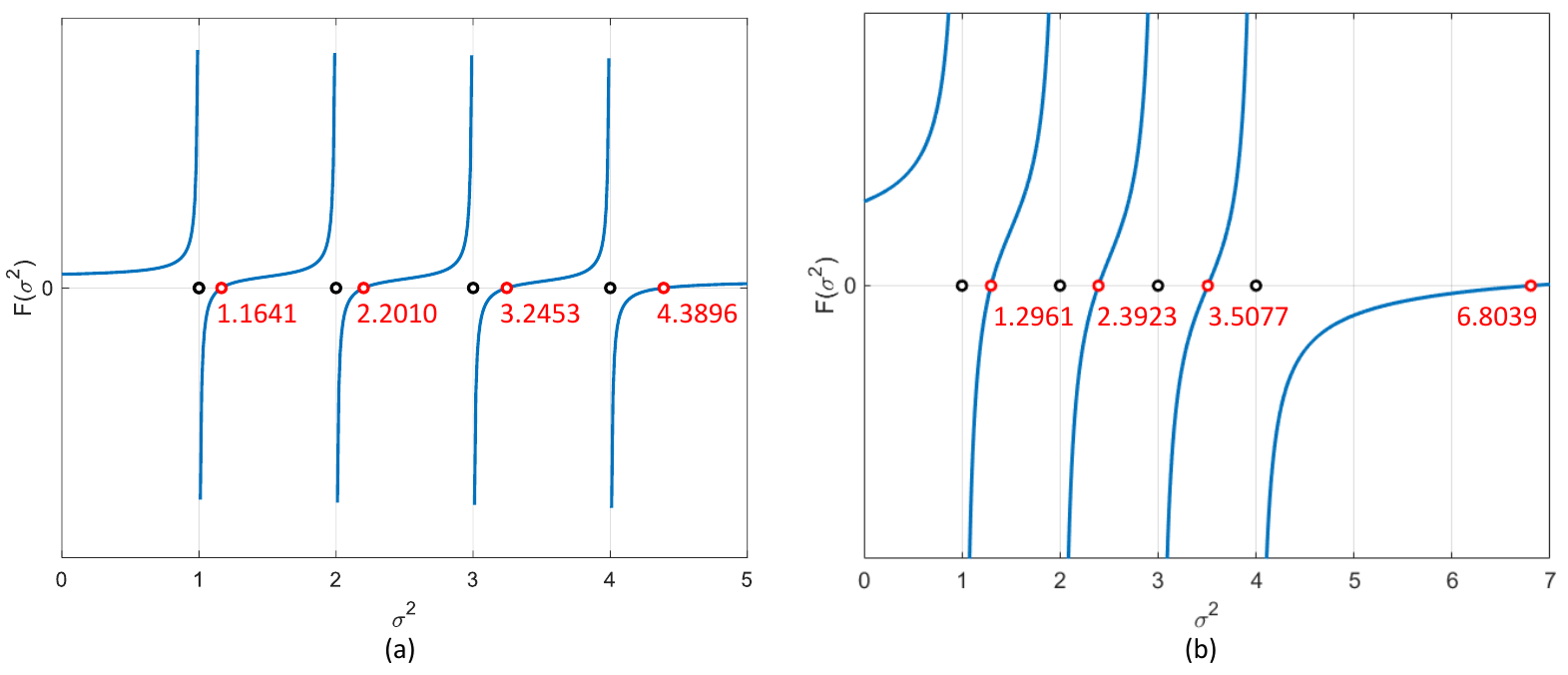}
\vspace{-0.2cm}
\caption{Graph of the secular equation with $\sigma_{i}^2(\qH_1) = (1,2,3,4)$ for (a) $ \| \qh' \|_2^2 = 1$ and (b) $ \| \qh' \|_2^2 = 4$.}
\label{fig:SingularValues_Examples}
\end{figure*}

The task of updating the SVD involves matrix perturbation processes. The following lemma characterizes the singular values of $\qH$.

\begin{Lemma}{\cite[Corollary 4.3.3]{HJ-Book}} \label{lemma:sv_increase}
The singular values of $\qH$ satisfy
\begin{equation} \label{eq:sigma h1 < h}
\sigma_{m}(\qH) \geq \sigma_{m}(\qH_1).
\end{equation}
\end{Lemma}

With Lemma \ref{lemma:sv_increase}, it becomes evident that the additional channel from $\qy_2$ definitively increases the \emph{effective channel gain} (or the singular values of $\qH$) of the system. To further elucidate the impact on the singular values of $\qH$ relative to $\qH_1$, we initially consider a special case where $N_2 = 1$. The derived results can be extended to general values of $N_2$ by iterating the obtained results. For $N_2 = 1$, we simply denote $\qH_2 \in \bbC^{1 \times M}$ by $\qh_2$ and $\qH'_2$ by $\qh' = [h'_1, \ldots, h'_M]$. We then express  \eqref{eq:A_def1} as:
\begin{equation}
 \qA =  \begin{bNiceMatrix}[margin]
        \Block[fill=red!10,rounded-corners]{3-3}{}
        \sigma_{1}(\qH_1)  & & \qzero \\
        &  \ddots & \\
        \qzero & & \sigma_{M}(\qH_1) \\ 
        \Block[fill=blue!10,rounded-corners]{*-*}{}
        h'_1 & \cdots & h'_M
       \end{bNiceMatrix} .
\end{equation}
Subsequently, \eqref{eq:AA_InUpdate} is formulated as:
\begin{equation} \label{eq:A_rankone}
 \qA^{\rm H} \qA = \qD_{{\rm H}_1}^2 + (\qh')^{ {\rm H}} \qh',
\end{equation}
representing a rank-one modification of $\qD_{{\rm H}_1}^2$.

Note that $\qh'= \qh_2 \qV_{{\rm H}_1}$ indicates that vector $\qh_2$ projects onto ${\qV_{{\rm H}_1} = [\qv_{{\rm H}_1,1}, \ldots, \qv_{{\rm H}_1,M}]}$, with each vector $\qv_{{\rm H}_1,m}$ corresponding to the right singular vectors of $\qH_1$ in decreasing order. The value of each element, $h'_i$, thus reflects the similarity between $\qh_2$ and each $\qv_{{\rm H}_1,i}$. For the rank-one modification of $\qD_{{\rm H}_1}^2$, we consider the following lemma:
\begin{Lemma} \label{lemma:secular_eq} \cite[Theorem 8.5.3]{Matrix-Book}
The singular values $\{ \sigma_{i}(\qH) \}$ satisfy the interlacing property
\begin{multline} \label{eq:interlacing}
    \sigma_{1}(\qH_1) + \| \qh' \|_2 \geq \sigma_{1}(\qH) \geq \sigma_{1}(\qH_1) > \cdots \\
    \geq \sigma_{M-1}(\qH_1) \geq \sigma_{M}(\qH) \geq \sigma_{M}(\qH_1) \geq 0,
\end{multline}
and the secular equation
\begin{equation} \label{eq:secular}
 F(\sigma^2) = 1 + \sum_{i=1}^{M} \frac{|h'_i|^2}{ \sigma_{i}^2(\qH_1) - \sigma^2 }  = 0.
\end{equation}
\end{Lemma}

Some insights can be observed from this lemma:

{\bf O1:} First, the secular equation indicates that $\sigma_{i}(\qH)$ should be larger than $\sigma_{i}(\qH_1)$ and follows the interlacing property \eqref{eq:interlacing}. Fig.~\ref{fig:SingularValues_Examples}(a) shows an example of the secular equation for $M=4$, $\sigma_{i}^2(\qH_1) = (1,2,3,4)$ and $\qh' = (1/2,1/2,1/2,1/2)$. The singular values of $\qH_1$ are marked in black ``$\circ$'', while those of $\qH$ are marked in red ``${\rl \circ}$''. Specifically, $\sigma_i^2(\qH)$ satisfies:
\begin{equation}
   1 + \frac{1}{4} \left( \frac{1}{ 4 - \sigma^2 } + \frac{1}{ 3 - \sigma^2 } + \frac{1}{ 2 - \sigma^2 } + \frac{1}{ 1 - \sigma^2 } \right) = 0.
\end{equation}
The secular function $F(\sigma^2)$ is monotone in between its poles. Thus, $F(\sigma^2)$ has precisely 4 roots, one in each of the intervals
\begin{equation}
   (1, 2), \, (1, 2), \, (3, 4), \, (4, \infty).
\end{equation}

{\bf O2:} Second, all the singular values undergo specific modifications from their corresponding $\sigma_{i}^2(\qH_1)$, with the largest singular value $\sigma_{\max}^2(\qH)$ experiencing the most substantial change and the smallest singular value $\sigma_{\min}^2(\qH)$ the least. To observe this, we let $\Delta_i = \sum_{m \neq i }  {|h'_m|^2}/{ (\sigma_{m}^2(\qH_1) - \sigma^2) }$. Applying \eqref{eq:secular}, we derive
\begin{equation} \label{eq:sigma_change}
    \sigma_i^2(\qH) = \sigma_{i}^2(\qH_1) + \frac{|h'_i|^2}{1+\Delta_i}.
\end{equation}
Note that $ \Delta_i$ has a lower bound of $-1$ but no upper bound. For the largest singular value, $\Delta_i$ assumes the most negative value; conversely, for the smallest singular value, it assumes the largest positive value, leading to our second observation.

{\bf O3:} Third, if $h'_i$ is small, then the change in $\sigma_{i}^2(\qH)$ from $\sigma_{i}^2(\qH_1)$ is also minimal. Conversely, if $h'_i$ dominates, the change in $\sigma_{i}^2(\qH)$ also dominates. This observation is directly derived from \eqref{eq:sigma_change}. For the special case where $h'_i = 0$, the singular value $\sigma_{i}^2(\qH)$ is unchanged.

{\bf O4:} Fourth, if $\qh' $ is uniformly spread across elements and $\| \qh' \|_2 $ sufficiently large, it will predominantly contribute to the largest singular value. This observation can be derived from the interlacing property  \eqref{eq:interlacing}. Fig.~\ref{fig:SingularValues_Examples}(b) demonstrates this observation.

With these observations, we conclude that by properly designing $\qh'$ (or equivalently $\qh_2$), we can improve the system's performance solely in the direction of increase. This capability of UE-CoMIMO is reminiscent of the recently popular technology of RIS. Like UE-CoMIMO, the additional path contributed by RIS can be viewed as a rank modification of $\qH_1$, expressed as
\begin{equation} \label{eq:RIS_rankone_update}
\qH = \qH_1 + \qa \qh'.
\end{equation}
However, unlike UE-CoMIMO, RIS does not necessarily enhance the singular values of the new channel $\qH$, as described in \eqref{eq:sigma h1 < h} in Lemma \ref{lemma:sv_increase}. As shown in \cite{Zhu2019}, the singular values $\sigma_{m}(\qH)$ of \eqref{eq:RIS_rankone_update} may be lower-bounded by $\sigma_{m}(\qH_1) - \|\qa\|_2 \| \qh'\|_2$. This indicates that if not properly designed, RIS can potentially reduce the effective channel gain.

\subsection{Customize $\qH_2$ Using PSs}

\begin{figure}
\centering
\includegraphics[width=3.0in]{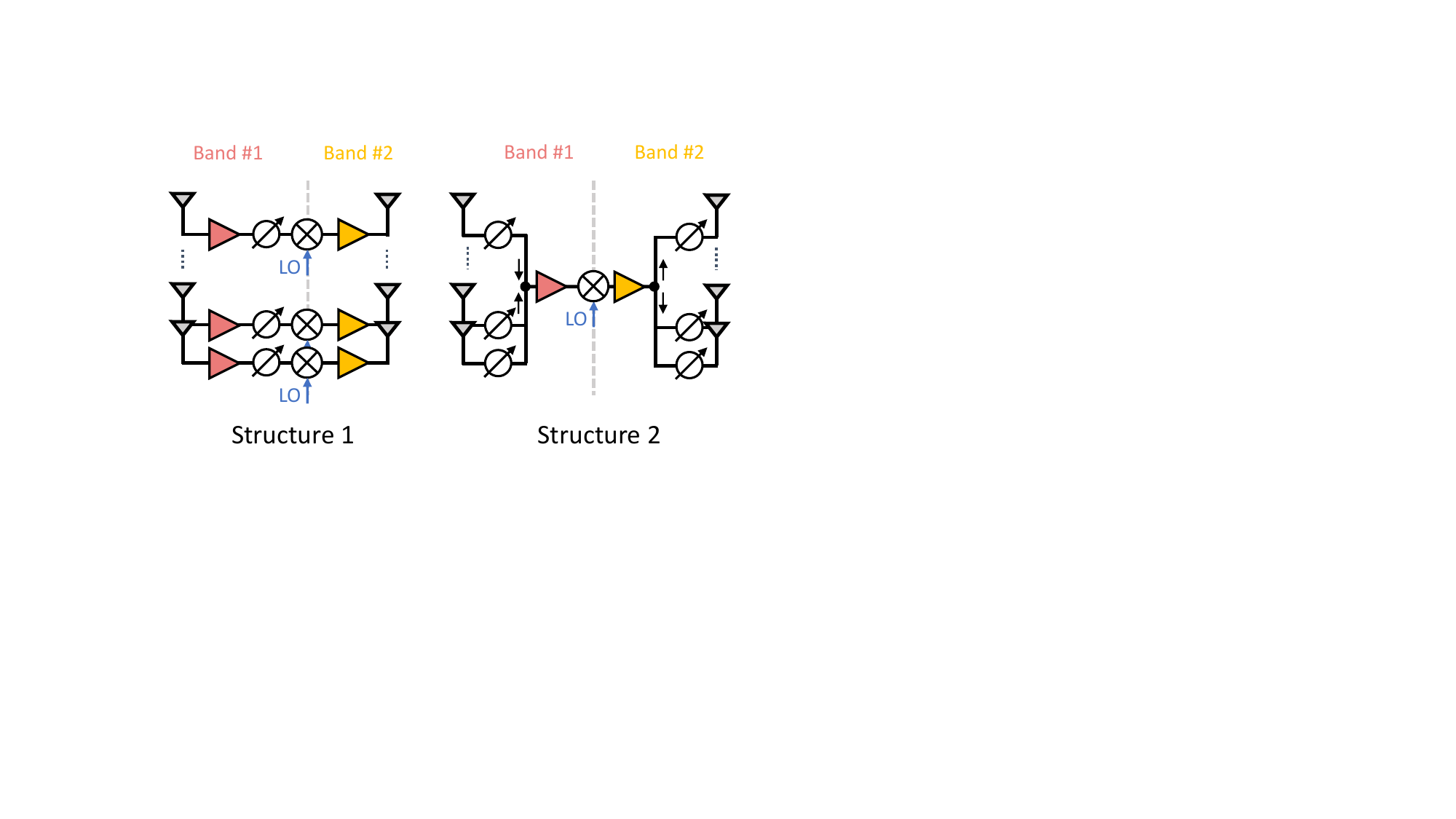}
\vspace{-0.2cm}
\caption{Illustration of two structures employing PSs at Co-UE.}
\label{fig:BF_structure}
\end{figure}

In the previous subsection, we demonstrated that the additional channel $\qH_2$, appended as a row to $\qH_1$, definitively increases the effective channel gain of the system. In this subsection, we aim to customize $\qH_2$ by introducing PSs. Two structures are considered for this purpose.

In {\bf Structure 1}, as shown in Fig.~\ref{fig:BF_structure}, each frequency-translation element consists of a receive antenna, a low-frequency LNA, a low-frequency PS, a high-frequency LNA, and a transmit antenna. With $N_{\rm c}$ antennas at Co-UE, there are $N_{\rm c}$ frequency-translation elements. For simplicity, we define the transmission coefficients of the PSs as $\qPhi_{\rm s1} = \diag( \qphi_{\rm s1}  )$ with $\qphi_{\rm s1} = [e^{j \theta_1}, \ldots, e^{j \theta_{N_{\rm c}}}]^{\rm T}$. Then, $\qH_2$ in Structure 1 is expressed as
\begin{equation} \label{eq:H2_s1}
    \qH_2 = \sqrt{\rho} \, \qH_{\rm p} \qPhi_{\rm s1}  \qH_{\rm c}  .
\end{equation}
Customization of $\qH_2$ can be achieved by adjusting $\theta_n \in (0,2\pi]$ for $n = 1, \ldots, N_{\rm c} $. Each element of $\qphi_{\rm s1}$ satisfies $|\phi_{{\rm s1},n}| = 1$, which we denote as $|\qphi_{\rm s1}| = \qone$.

In {\bf Structure 2}, as shown in Fig.~\ref{fig:BF_structure}, each frequency-translation element includes $N_{\rm c}$ receive antennas, $N_{\rm c}$ low-frequency PSs, a power combiner, a low-frequency LNA, a mixer, a high-frequency LNA, a power splitter, $N_{\rm c}$ high-frequency PSs, and $N_{\rm c}$ transmit antennas. The coefficients for the PSs used in the power combiner and power splitter are defined as
\begin{subequations}
\begin{align}
 \qphi_{\rm r} &= [ e^{j \theta_{{\rm r},1}}, \ldots, e^{j \theta_{{\rm r},N_{\rm c}}}  ]^{\rm T}, \\
 \qphi_{\rm t} &= [ e^{j \theta_{{\rm t},1}}, \ldots, e^{j \theta_{{\rm t},N_{\rm c}}}  ]^{\rm T}.
\end{align}
\end{subequations}
The channel $\qH_2$ in Structure 2 is expressed as
\begin{equation} \label{eq:H2_s2}
    \qH_2 = \sqrt{\rho} \, \qH_{\rm p} \qphi_{\rm t} \qphi_{\rm r}^{\rm H}  \qH_{\rm c}  = \sqrt{\rho} \, \qH_{\rm p} \qPhi_{\rm s2} \qH_{\rm c},
\end{equation}
where $\qPhi_{\rm s2} = \qphi_{\rm t} \qphi_{\rm r}^{\rm H} $. Customization of $\qH_2$ is achieved by adjusting $( \theta_{{\rm r},n}, \theta_{{\rm t},n})$ for $n = 1, \ldots, N_{\rm c} $. Both vectors satisfy $|\qphi_{\rm t}| = \qone$ and $|\qphi_{\rm r}| = \qone$.

\begin{table}[!t]
\caption{Power Consumption of RF Elements.}\label{tab:Power_RFElement}
\centering
\begin{tabular}{|l|l|r|}
\hline
 \textbf{Element} & \textbf{Product \#} & \textbf{Power} \\
\hline\hline
 Low Frequency LNA & ZX60-3800LN-S+ & 425 mW \\
\hline
 Low Frequency PS & MAP-010144 & 0.15 mW \\
\hline
 High Frequency LNA & ZX60-05113LN+ & 180 mW \\
\hline
 High Frequency PS & TGP2105-SM & 0.15 mW \\
\hline
 Mixer & ZX05-83-S+ & $\sim$ 0 mW \\
\hline
\end{tabular}
\end{table}

For ease of expression, we use $\qPhi_{\rm s}$ to represent either $\qPhi_{\rm s1}$ or $\qPhi_{\rm s2}$. In both structures, the channel $\qH_2$ is determined by the product of three matrices: $\qH_{\rm p}$, $\qPhi_{\rm s}$, and $\qH_{\rm c}$. The rank of $\qH_2$ is determined by:
\[ \rank(\qH_2) =  \min\{ \rank(\qH_{\rm p}), \rank(\qPhi_{\rm s}), \rank(\qH_{\rm c}) \}. \]
In Structure 2, $\qH_2$ is rank-1, leading to a rank-1 modification of $\qH_1$. As noted in the observations from Lemma \ref{lemma:secular_eq} (specifically {\bf O2} and {\bf O3}), a rank-1 modification primarily impacts the largest singular value of $\qH_1$, resulting in limited overall changes to the other singular values. In contrast, Structure 1 enables a rank-$N_2$ modification of $\qH_1$, allowing for a more comprehensive update to its singular values. This capability makes Structure 1 more versatile in enhancing the system's SE. However, this added flexibility comes at a cost: Structure 1 requires multiple mixers, leading to higher implementation complexity and increased power consumption compared to Structure 2.

In Section \ref{sec:Experiments}, we implement both structures. Table \ref{tab:Power_RFElement} displays the power consumption of the RF elements involved. For each structure, the total power consumption is calculated as
\begin{subequations} \label{eq:power_consumption}
\begin{align}
 P_{{\rm s1}} &= N_{\rm c} \times (425 +0.15+180), \\
 P_{{\rm s2}} &= N_{\rm c} \times 0.15 + 425 + 180 + N_{\rm c} \times 0.15 .
\end{align}
\end{subequations}
The power consumption of the PSs is negligible compared to that of the LNAs. Consequently, the power consumption of Structure 1 is approximately $N_{\rm c}$ times higher than that of Structure 2. For example, with $N_{\rm c} = 4$, the calculated power consumption is $P_{{\rm s1}} = 2420.6$ mW and $P_{{\rm s2}} = 606.2$ mW.

\subsection{Maximizing SE}
To specify a method for adjusting $\qPhi_{\rm s}$, consider the SE of the UE-CoMIMO system, which is given by:
\begin{align}
 C &= \log \det {\left(\qI_{N_1+N_2} + \qH\qH^{\rm H}\right)},   \label{eq:Cap1} \\
   &= \log \det {\left( \qI_{M} +  \qD_{{\rm H}_1}^2 + (\qH'_2)^{ {\rm H}} \qH'_2  \right)}, \notag \\
   &= \underbrace{\log \det {\left( \qI_{M} +  \qD_{{\rm H}_1}^2 \right)}}_{\text{SE of original Pr-UE}} \notag \\
   & \quad + \underbrace{\log \det{\left(  \qI_{N_2} + \qH'_2 (\qI_{M} +  \qD_{{\rm H}_1}^2)^{-1} (\qH'_2)^{ {\rm H}} \right)}}_{\text{Gain from Co-UE}}, \label{eq:Cap2}
\end{align}
where the second equality follows from footnote 1 and \eqref{eq:AA_InUpdate_2}, and in the third equality, we utilize the fact that $\det(\qA\qB) = \det(\qA) \det(\qB)$ and $\det(\qI+\qA\qB) = \det(\qI+\qB\qA)$.
The second term of \eqref{eq:Cap2} represents the contribution from the additional channel $\qy_2$ of the UE-CoMIMO system, which heavily depends on $\qH'_2$.
Recall from the previous subsection that the modification of singular values also heavily depends on $\qH'_2$.

We first consider maximizing SE with Structure 2. Recall that $\qH'_2 = \qH_2 \qV_{{\rm H}_1}$. Substituting \eqref{eq:H2_s2}, we have
\begin{equation}
    \qH'_2  = \qH_2 \qV_{{\rm H}_1} = \sqrt{\rho} \, \qH_{\rm p} \qphi_{\rm t} \qphi_{\rm r}^{\rm H} \qH_{\rm c} \qV_{{\rm H}_1}.
\end{equation}
The Gram matrix $(\qH'_2)^{\rm H}  \qH'_2 $ is expressed as
\begin{equation} \label{eq:Gram_H2}
    (\qH'_2)^{\rm H}  \qH'_2 = \rho \,  \qV_{{\rm H}_1}^{\rm H} \qH_{\rm c}^{\rm H} \qphi_{\rm r} \qphi_{\rm t}^{\rm H} \qH_{\rm p}^{\rm H} \qH_{\rm p} \qphi_{\rm t} \qphi_{\rm r}^{\rm H} \qH_{\rm c} \qV_{{\rm H}_1}.
\end{equation}
Defining $ \eta = \qphi_{\rm t}^{\rm H} \qH_{\rm p}^{\rm H} \qH_{\rm p} \qphi_{\rm t} $ and $\qomega = \qV_{{\rm H}_1}^{\rm H} \qH_{\rm c}^{\rm H} \qphi_{\rm r} $, and substituting them into \eqref{eq:Cap2}, the SE becomes
\begin{equation} \label{eq:Cap3}
 C =  \sum_{m=1}^{M} \log(1+\sigma_{m}^2(\qH_1)) +  \log{\left( 1+ \rho \,  \eta \, \qomega^{\rm H} (\qI +  \qD_{{\rm H}_1}^2)^{-1}  \qomega \right)}.
\end{equation}
Maximizing the SE involves maximizing $ \rho \,  \eta \, \qomega^{\rm H} (\qI +  \qD_{{\rm H}_1}^2)^{-1}  \qomega = \rho \,  \eta \, \qphi_{\rm r}^{\rm H} \qH_{\rm c} \qV_{{\rm H}_1} (\qI +  \qD_{{\rm H}_1}^2)^{-1} \qV_{{\rm H}_1}^{\rm H} \qH_{\rm c}^{\rm H} \qphi_{\rm r}$, leading to the optimization problem:
\begin{align}
    \text{(P2):} & & \max_{\qphi_{\rm t},\, \qphi_{\rm r}} &  ~~ \rho \,  \eta \, \qphi_{\rm r}^{\rm H} \qH_{\rm c} (\qI_{M}+\qH_1^{\rm H} \qH_1 )^{-1} \qH_{\rm c}^{\rm H} \qphi_{\rm r}, \notag \\
    & & {\rm s.t.} &~~  |\qphi_{\rm t} |  = \qone, ~  |\qphi_{\rm r} |  = \qone. \notag 
\end{align}

Interestingly, the optimality of $\qphi_{\rm t}$ can be achieved solely by maximizing $\eta$. Consequently, the optimality of $\qphi_{\rm t}$ and $\qphi_{\rm r}$ can be determined separately. This characteristic significantly simplifies the complexity involved in determining their optimal configurations. Define
\begin{equation}
 \widetilde{\qH}_{\rm c} = \qH_{\rm c} \left(\qI_{M}+\qH_1^{\rm H} \qH_1 \right)^{-1/2},
\end{equation}
and let
\begin{equation}
 \qOmega_{\rm t} = \qphi_{\rm t} \qphi_{\rm t}^{\rm H},~~ \qOmega_{\rm r} = \qphi_{\rm r} \qphi_{\rm r}^{\rm H}.
\end{equation}
Then, the optimality of $\qphi_{\rm t}$ and $\qphi_{\rm r}$ can be formulated into the following two problems:
\begin{align*}
    \text{(P2-1):} & & \max_{\qOmega_{\rm t}} &  ~~ \tr( \qH_{\rm p}^{\rm H} \qH_{\rm p} \qOmega_{\rm t}),  \\
    & & {\rm s.t.} &~~  [\qOmega_{\rm t}]_{n,n} = 1,~n=1,\ldots, N_{\rm c},
    ~ \qOmega_{\rm t} \succeq \qzero   
\end{align*}
and
\begin{align*}
    \text{(P2-2):} & & \max_{\qOmega_{\rm r}} &  ~~ \tr( \widetilde{\qH}_{\rm c} \widetilde{\qH}_{\rm c}^{\rm H} \qOmega_{\rm r}), \\
    & & {\rm s.t.} &~~  [\qOmega_{\rm r}]_{n,n} = 1,~n=1,\ldots, N_{\rm c},
    ~ \qOmega_{\rm r} \succeq \qzero. 
\end{align*}
These problems also need to satisfy the rank constraints: $\rank(\qOmega_{\rm t}) = 1$ and $\rank(\qOmega_{\rm r}) = 1$, consistent with the definitions of $\qOmega_{\rm t}$ and $\qOmega_{\rm r}$. Together with the above discussions, we present the following theorem.

\begin{Theorem} \label{theorem:sep_opt&special_case}
For Structure 2, the optimality of $\qphi_{\rm t}$ and $\qphi_{\rm r}$ can be determined separately.
In a special case where $\rank(\qH_{\rm p}) = 1$ and $\rank(\widetilde{\qH}_{\rm c}) = 1$, the maximal SE can be achieved by setting
\begin{equation} \label{eq:phi_opt}
 \qphi_{\rm t} = e^{-j \arg(\qv_{{\rm H}_{\rm p},1})},  ~~ \qphi_{\rm r} = e^{-j \arg(\qu_{\widetilde{\rm H}_{\rm c},1})},
\end{equation}
where $\qv_{{\rm H}_{\rm p},1}$ is the first right singular vector of $\qH_{\rm p}$, and $\qu_{\widetilde{\rm H}_{\rm c},1}$ is the first left singular vector of $\widetilde{\qH}_{\rm c}$.
\end{Theorem}

Later in Section \ref{sec:Ray-Tracing-Indoor}, we demonstrate that the case where $\rank(\qH_{\rm p}) = 1$ frequently arises in practical applications as the distance between the Co-UE and Pr-UE increases. However, the two optimization problems in (P2-1) and (P2-2) are generally non-convex and challenging to solve. To address this, we may employ semidefinite relaxation (SDR) to relax the rank-one constraint. Nevertheless, the relaxed problem may not yield a rank-one solution, necessitating additional steps to derive a rank-one solution from the higher-rank result. In this paper, we adopt a widely used approach \cite{Wu-18GLOBECOM}: generating a set of complex Gaussian random vectors with zero mean and covariance matrices corresponding to $\qOmega_{\rm t}$ and $\qOmega_{\rm r}$. These random vectors are substituted into \eqref{eq:phi_opt} to produce a set of candidate $\qphi_{\rm t}$ and $\qphi_{\rm r}$ solutions. The solution that achieves the maximum SE among the set is selected as the final solution.

Next, we consider maximizing the SE for Structure 1. Applying \eqref{eq:H2_s1}, we have
\begin{equation} \label{eq:H'_2wS2}
    \qH'_2  = \qH_2 \qV_{{\rm H}_1} = \sqrt{\rho} \, \qH_{\rm p} \qPhi_{\rm s1} \qH_{\rm c} \qV_{{\rm H}_1}.
\end{equation}
Moreover, $\widetilde{\qH}_{\rm c} \widetilde{\qH}_{\rm c}^{\rm H}$ can be expressed as
\begin{equation}
 \widetilde{\qH}_{\rm c} \widetilde{\qH}_{\rm c}^{\rm H}  = \sum_{\ell=1}^{{\sf rank}(\qH_{\rm c})} \sigma_{{\widetilde{\rm H}_{\rm c}},\ell}^2 \qu_{{\widetilde{\rm H}_{\rm c}},\ell} \qu_{{\widetilde{\rm H}_{\rm c}},\ell}^{\rm H},
\end{equation}
Thus, the Gram matrix $\qH'_2 (\qI_{M} +  \qD_{{\rm H}_1}^2)^{-1} (\qH'_2)^{ {\rm H}} $ is expressed as
\begin{align}
    & \qH'_2 (\qI_{M} +  \qD_{{\rm H}_1}^2)^{-1} (\qH'_2)^{ {\rm H}} \notag \\
    &= \rho \, \qH_{\rm p} \qPhi_{\rm s1} \qH_{\rm c} \qV_{{\rm H}_1} (\qI_{M} +  \qD_{{\rm H}_1}^2)^{-1} \qV_{{\rm H}_1}^{\rm H} \qH_{\rm c}^{\rm H} \qPhi_{\rm s1}^{\rm H} \qH_{\rm p}^{\rm H}, \notag \\
    &= \rho \, \sum_{\ell=1}^{{\sf rank}(\qH_{\rm c})} \sigma^2_{{\widetilde{\rm H}_{\rm c}},\ell} \qH_{\rm p} \qPhi_{\rm s1}  \qu_{{\widetilde{\rm H}_{\rm c}},\ell} \qu_{{\widetilde{\rm H}_{\rm c}},\ell}^{\rm H}  \qPhi_{\rm s1}^{\rm H} \qH_{\rm p}^{\rm H}, \notag \\
    &= \rho \, \sum_{\ell=1}^{{\sf rank}(\qH_{\rm c})} \widetilde{\qH}_{{\rm p},\ell} \qphi_{\rm s1} \qphi_{\rm s1}^{\rm H} \widetilde{\qH}_{{\rm p},\ell}^{\rm H},
\end{align}
where $\widetilde{\qH}_{{\rm p},\ell} = \sigma_{{\widetilde{\rm H}_{\rm c}},\ell} \qH_{\rm p} \diag(\qu_{{\widetilde{\rm H}_{\rm c}},\ell}  )$. Define $\qOmega_{\rm s1} = \qphi_{\rm s1} \qphi_{\rm s1}^{\rm H}$.
Consequently, maximizing the SE in \eqref{eq:Cap2} can be equivalently expressed as
\begin{align*}
     \text{(P1):}& & \max_{\qphi_{\rm s1}} &~~ \log\det {\left( \qI_{N_2} + \rho \, \sum_{\ell=1}^{{\sf rank}(\qH_{\rm c})} \widetilde{\qH}_{{\rm p},\ell} \qOmega_{\rm s1} \widetilde{\qH}_{{\rm p},\ell}^{\rm H}  \right)},  \\
     & & {\rm s.t.} &~~ [\qOmega_{\rm s1}]_{n,n} = 1,~n=1,\ldots, N_{\rm c},
    ~ \qOmega_{\rm s1} \succeq \qzero, 
\end{align*}
where $\qOmega_{\rm s1}$ must also satisfy the rank-one constraint.
 
\begin{Theorem} \label{theorem:special_case_S2}
For Structure 1, in a special case where $\rank(\qH_{\rm p}) = 1$ and $\rank(\widetilde{\qH}_{\rm c}) = 1$, the maximal SE can be achieved by setting
\begin{equation}
 \qphi_{\rm s1} = e^{-j \arg( \qv_{{\rm H}_{\rm p},1} \odot \qu_{\widetilde{\rm H}_{\rm c},1})}.
\end{equation}
\end{Theorem}

Theorem \ref{theorem:special_case_S2} demonstrates that the optimal $\qphi_{\rm s1}$ in Structure 1 should simultaneously compensate for the channel phases of the first hop from the BS to the Co-UE ($\widetilde{\qH}{\rm c}$) and the second hop from the Co-UE to the Pr-UE ($\qH_{\rm p}$). This strategy contrasts with Structure 2, where $\qphi_{\rm t}$ and $\qphi_{\rm r}$ are used to separately compensate $\qH_{\rm p}$ and $\widetilde{\qH}_{\rm c}$, respectively.

For general cases, an analytical solution to (P1) is not available. Similar to (P2), the solution can be numerically approximated using the SDR approach, followed by additional steps to derive a rank-one solution \cite{Zhang-20JSAC}. Problem (P1) reveals that the optimal $\qphi_{\rm s1}$ in Structure 1 depends on multiple $\widetilde{\qH}_{{\rm p},\ell}$ for $\ell = 1, \dots ,{\sf rank}(\qH_{\rm c})$. Given that $\widetilde{\qH}_{{\rm p},\ell} = \sigma_{{\widetilde{\rm H}_{\rm c}},\ell} \qH_{\rm p} \diag(\qu_{{\widetilde{\rm H}_{\rm c}},\ell}  ) $, it can be inferred that as $\qH_{\rm p}$ becomes more isotropic, simultaneously compensating for multiple phases becomes increasingly challenging. Consequently, in such cases, the optimality of $\qphi_{\rm s1}$ is expected to yield only marginal gains compared to the case without phase adjustment, i.e., $\qphi_{\rm s1} = \qone$.

\subsection{Practical Algorithm}

Solving (P1) and (P2) relies on full CSI for $(\qH_1, \qH_{\rm c}, \qH_{\rm p})$. However, estimating the individual channels of the Co-UE link, i.e., $\qH_{\rm c}$ and $\qH_{\rm p}$, is highly challenging.\footnote{If Co-UE is capable of demodulating the signal from $\qy_{\rm c}$, the channel matrix $\qH_{\rm c}$ can be estimated \cite{Chen-24}. However, this increases the complexity and computational load of Co-UE.} Additionally, the solutions derived from (P1) and (P2) yield continuous phase values, which require infinite phase resolution. In practical implementations, discrete and finite phase resolutions are more feasible.

To address these challenges, we adopt a blind method known as the Blind Greedy (BG) algorithm \cite{Chian-24TAP}, which is based on the greedy algorithm. The BG algorithm consists of two main steps: Random-Max Sampling (RMS) and Greedy Searching (GS):
\begin{itemize}
\item {\bf RMS}: This step randomly samples from a broad set of phase combinations to identify a good initial point for the greedy algorithm. After a few trials, the phase states that yield the maximum SE in \eqref{eq:Cap1} are selected as the starting point for the GS step.

\item {\bf GS}: Starting from the initial PS state, this step optimizes each PS sequentially by exhausting the states that yield the maximum SE for one PS at a time. The best state is selected, and the PS is updated accordingly. This process continues iteratively for each PS until all are optimized. A complete update of all PSs is referred to as a ``round,'' and multiple rounds may be performed to further refine the states for improved SE.
\end{itemize}

To calculate the SE in \eqref{eq:Cap1}, the channels $\qH_1$ and $\qH_2$ are required, where $\qH_2$ represents the compound channel rather than the individual channels $\qH_{\rm c}$ and $\qH_{\rm p}$. Both $\qH_1$ and $\qH_2$ can be estimated at Pr-UE. Consequently, Pr-UE can execute the BG algorithm based on its measured SE. In Section \ref{sec:Device&Implementation}, we provide a detailed implementation of the BG algorithm.

For Structure 1, the BG algorithm can be applied directly. For Structure 2, since the optimality of $\qphi_{\rm t}$ and $\qphi_{\rm r}$ can be determined separately, the BG algorithm is applied sequentially to $\qphi_{\rm r}$ and $\qphi_{\rm t}$. Specifically, $\qphi_{\rm t}$ is adjusted to optimally transmit towards Pr-UE, while $\qphi_{\rm r}$ is tuned to align with the incoming channel from the BS. This alternate adjustment of $\qphi_{\rm t}$ and $\qphi_{\rm r}$ is repeated in subsequent iterations. In Section \ref{sec:Optimality_S2}, we discuss whether $\qphi_{\rm t}$ or $\qphi_{\rm r}$ should be adjusted first. Additionally, in the simulation section, we demonstrate that the BG algorithm is highly effective, achieving SE values that closely approximate those obtained by solving (P1) and (P2).

\section{Simulations and Discussions}
\label{sec:Sim&Dis}

\subsection{Optimality Properties of $\qphi_{\rm t}$ and $\qphi_{\rm r}$} \label{sec:Optimality_S2}

\definecolor{mypink1}{rgb}{0, 0, 1}
\definecolor{mypink2}{rgb}{0, 1, 0}
\definecolor{mypink3}{rgb}{0, 1, 1}

\begin{figure*}
\centering
\includegraphics[width=7.00in]{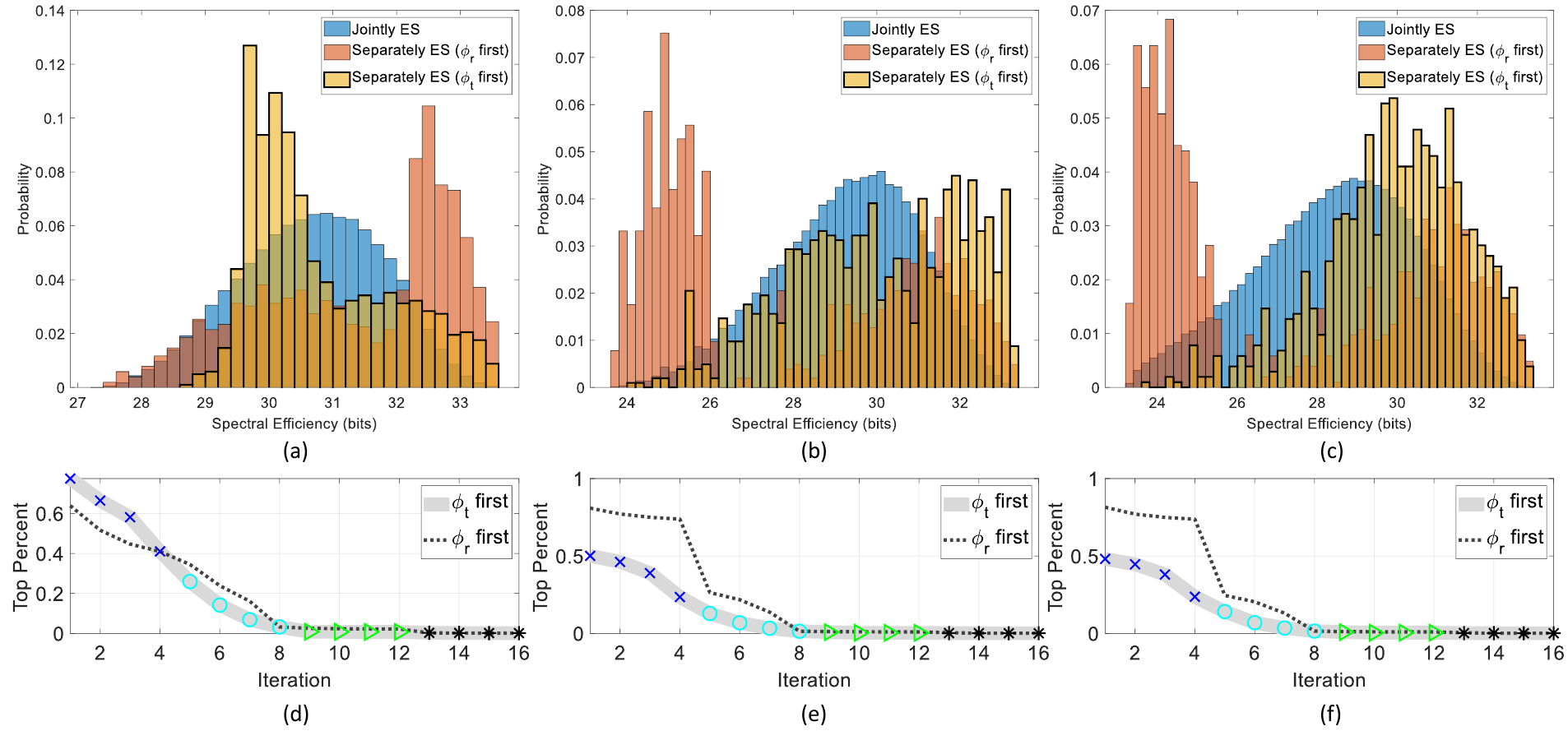}
\caption{Histograms of the SE achieved by joint ES and separate ES in three different positions: (a) $(1,1,1),\lambda_{\rm H}$, (b) $(5,5,5),\lambda_{\rm H}$, and (c) $(10,10,10),\lambda_{\rm H}$. (d)--(f) depict the corresponding trajectories of the BG algorithm. The achieved SEs are marked by \textcolor{mypink1}{$\times$}, \textcolor{mypink2}{$\circ$}, \textcolor{mypink3}{$\triangleright$}, and $\ast$ for the iterations $[1 : N_{\rm c}]$, $[N_{\rm c}+1 : 2N_{\rm c}]$, $[2N_{\rm c}+1 : 3N_{\rm c}]$, and $[3N_{\rm c}+1 : 4N_{\rm c}]$.}
\label{fig:S2_Hist}
\end{figure*}

This subsection investigates the optimality properties of $\qphi_{\rm t}$ and $\qphi_{\rm r}$ for Structure 2. For this study, we set $M = N_1 = N_2 = N_{\rm c} = 4$. The statistical Rician fading channel model is used to simplify the channel analysis, with a more realistic ray-tracing model discussed later. The Rician fading channel matrices are generated as
\begin{equation} \label{eq:H_ast}
 \qH_{\ast} = \sqrt{\frac{1}{\kappa_{\ast}+1}} \qH_{{\rm iid},\ast} + \sqrt{\frac{\kappa_{\ast}}{\kappa_{\ast}+1}} \qH_{{\rm LOS},\ast} ,
\end{equation}
where ${\ast} \in \{ 1, {\rm c}, {\rm p} \}$. For simplicity, ``${\ast}$'' is often omitted when the descriptions are applicable to all cases.
Here, $\qH_{\rm iid}$ represents the NLOS component with entries following an i.i.d. standard complex Gaussian distribution, while $\qH_{\rm LOS}$ represents the LOS component with entries modeled as \cite{Eric-TCOM11}:
\begin{equation} \label{eq: H_LOS}
 [\qH_{\rm LOS}]_{n,m} = e^{-j2\pi \frac{d_{n,m}}{\lambda_{{\rm L}/{\rm H}}}},
\end{equation}
where $\lambda_{{\rm L}/{\rm H}}$ is the wavelength corresponding to the $f_{\rm L}$ or $f_{\rm H}$ band, and $d_{n,m}$ denotes the distance between transmit antenna $m$ and receive antenna $n$. The $\kappa$-factor determines the gain division between the LOS and NLOS parts. We use the SNR to reflect signal quality and, for conciseness, omit the large-scale fading effect.\footnote{The large-scale fading effect, also known as free-space path loss, can be incorporated by introducing the factor ${\lambda_{{\rm L}/{\rm H}}^2}/{(4\pi d_{\ast})^2}$ into \eqref{eq:H_ast}, where $d_{\ast}$ represents the distance between the transmitter and receiver. Accounting for large-scale fading would make the SNR dependent on the relative positions of Co-UE and Pr-UE. Alternatively, the SNR can be adjusted by varying the noise level, bypassing explicit modeling of free-space path loss.}

In this simulation, Pr-UE is positioned at the origin, while the Co-UE is located at three different positions relative to Pr-UE: $(1,1,1) \,\lambda_{\rm H}$, $(5,5,5) \,\lambda_{\rm H}$, and $(10,10,10) \,\lambda_{\rm H}$. These locations are normalized with respect to their operational wavelengths. Variations in the Co-UE positions cause the channel matrix $\qH_{{\rm LOS},{\rm p}}$ to transition from full rank (near-field) to rank-1 (far-field). The $\kappa$-factors are set as $\kappa_{1} = \kappa_{\rm c} = 0$ and $\kappa_{\rm p} = \infty$. Consequently, $\qH_1$ and $\qH_{\rm c}$ consist entirely of i.i.d. Rayleigh fading channels, while $\qH_{\rm p}$ consists entirely of LOS components. This setup represents a scenario where the link between the BS and the Pr-UE is characterized by rich multipath propagation, while the Co-UE and the Pr-UE are in close proximity, resulting in their link being dominated by a LOS path. The SNR is set to 20 dB, measured based on channel $\qH_1$ and noise $\qz_1$ at the Pr-UE.

A 3-bit phase resolution (i.e., $Q=8$ levels) is used for each PS. Fig.~\ref{fig:S2_Hist} shows the histograms of SE values. The blue bins represent the SE distribution obtained through joint exhaustive search (ES), which evaluates all $Q^{N_{\rm c}} \times Q^{N_{\rm c}}$ combinations of $(\qphi_{\rm t}, \qphi_{\rm r})$. For instance, in Fig.~\ref{fig:S2_Hist}(a), the SE values are distributed between 27.2 and 33.6 bits/Hz, with the most frequent value occurring around 31 bits/Hz. The figure also displays the results of separate ES, which explores $2 \times Q^{N_{\rm c}}$ combinations by separately optimizing $\qphi_{\rm t}$ and $\qphi_{\rm r}$. Separate ES reduces complexity by evaluating only a subset of combinations considered in joint ES. Two strategies for separate ES are analyzed:
\begin{itemize}
\item {\bf $\qphi_{\rm t}$ first (yellow bins):} ES is performed first on $\qphi_{\rm t}$ and then on $\qphi_{\rm r}$.
\item {\bf $\qphi_{\rm r}$ first (red bins)}: ES is performed first on $\qphi_{\rm r}$ and then on $\qphi_{\rm t}$.
\end{itemize}
The figures show that, across all Co-UE positions, both joint ES and separate ES achieve the same optimal SE, validating Theorem \ref{theorem:sep_opt&special_case}. Additionally, the order of separate ES (i.e., $\qphi_{\rm t}$ first or $\qphi_{\rm r}$ first) does not affect the final result. However, if only a few random trials are available, $\qphi_{\rm r}$ first performs better in Fig.~\ref{fig:S2_Hist}(a), where its histogram covers a higher SE region than those of $\qphi_{\rm t}$ first. Conversely, $\qphi_{\rm t}$ first is preferred in Figs.~\ref{fig:S2_Hist}(b) and \ref{fig:S2_Hist}(c). {\bf This demonstrates that the preference for $\qphi_{\rm t}$ or $\qphi_{\rm r}$ depends on the channel characteristics.} For instance, in Fig.~\ref{fig:S2_Hist}(a), where $\qH_{\rm p}$ is full rank, $\qphi_{\rm t}$ first does not significantly enhance SE compared to the other scenarios.

Figs.~\ref{fig:S2_Hist}(d)--(f) present the averaged trajectory of the BG algorithm over 10,000 trials of random channels. The vertical axis indicates the percentile rank of SE achieved relative to all possible combinations explored in separate ES. The results demonstrate a significant improvement in rank with each BG iteration, ultimately converging to the optimal rank. Key observations include:
\begin{itemize}
\item After the first $N_{\rm c}$ iterations, the algorithm reaches the top 20\%, depending on the priority given to $\qphi_{\rm t}$ or $\qphi_{\rm r}$.
\item By the end of $2 N_{\rm c}$ iterations (regardless of the $\qphi_{\rm t}$ and $\qphi_{\rm r}$ order), the algorithm typically achieves the top 1\%.
\item Additional iterations provide only marginal improvements, helping $\qphi_{\rm t}$ and $\qphi_{\rm r}$ refine to nearly optimal values.
\end{itemize}
The trajectories for the $\qphi_{\rm t}$ first and $\qphi_{\rm r}$ first strategies further highlight {\bf the advantage of prioritizing $\qphi_{\rm t}$ first in scenarios where $\qH_{\rm p}$ is dominated by a single strong path (e.g., Figs.~\ref{fig:S2_Hist}(e) and \ref{fig:S2_Hist}(f)). Nonetheless, both strategies eventually converge to the same optimal SE.}

Table \ref{tab:Complexity} compares the number of trials required by the algorithms and provides an example. The BG algorithm requires $2 Q N_{\rm c} \times I$ trials to find the optimal PSs, where $I$ represents the number of rounds of $2 N_{\rm c}$ iterations. In our simulations, we typically set $I=2$ to ensure convergence, though the difference in SE performance between $I=1$ and $I=2$ is negligible. Clearly, the BG algorithm's trial time is significantly lower than that of the joint ES and separate ES methods.

Notably, the optimization problem in (P2) relies on channel knowledge, allowing it to bypass the iterative testing and probing required by the aforementioned search-based algorithms. In this paper, the SDR problems (P2-1) and (P2-2) are solved using {\tt CVX}, a software package for specifying and solving convex optimization problems \cite{cvx}. The worst-case complexity of SDR is approximately $N_{\rm c}^{4.5} \log(1/\epsilon)$, where $\epsilon > 0$ denotes the solution accuracy \cite{Luo-10-SPMag}. After solving the SDR, additional steps are required to extract a rank-one solution from the higher-rank results. To provide a comprehensive view of the computational burden, Table \ref{tab:Complexity} also includes the execution times for each algorithm.\footnote{Execution times were measured using MATLAB R2023b on an Intel(R) Core(TM) i7-8700 CPU @ 3.20GHz with 8.00 GB RAM.} The results clearly show that the complexity of obtaining optimal continuous phases using SDR is significantly higher than that of the BG algorithm.

\begin{table}[t]
\centering
\caption{Time complexity and Example for $Q =8$, $N_{\rm c}=4$, $I = 2$.} \label{tab:Complexity}
\begin{tabular}{|l|l|l|l|}
\hline
                   & Trials & Example Value & Times (ms) \\ \hline
\textbf{Joint ES}  & $Q^{N_{\rm c}} \times Q^{N_{\rm c}}$  & 16,777,216 & 6.5e4 \\ \hline
\textbf{Separate ES}  & $2 \times Q^{N_{\rm c}}$  & 8,192  & 31.7 \\ \hline
\textbf{BG}  & $2 Q N_{\rm c} I $  & 64 &  0.6 \\ \hline
{\bf Continuous Phase}  & \multirow{2}{*}{---}  & \multirow{2}{*}{---} &  \multirow{2}{*}{753.7} \\
{\bf via (P2-1)\&(P2-2)} & & & \\ \hline
\end{tabular}
\end{table}

\subsection{Advantage of UE-CoMIMO} \label{sec: sim_adv_UECoMIMO}
Next, we discuss the advantages of the UE-CoMIMO system between Structures 1 and 2. Unless otherwise specified, the scenarios and parameters are set as in the previous subsection.

We begin with Structure 2, analyzing phase resolutions at different levels: infinite, and $Q = 32$, $16$, and $2$ levels. For finite-resolution cases, we apply either ES or the BG algorithm to obtain the optimal solution of $(\qphi_{\rm t}, \qphi_{\rm r})$. For infinite phase resolution, we use SDP to solve problems (P2-1) and (P2-2), denoted as optimal (continuous) phases. We also examine the case where $\qphi_{\rm r}$ and $\qphi_{\rm t}$ allow amplitude adjustments. Here, the maximum SE is achieved by setting $\qphi_{\rm t} = \qv_{{\rm H}{\rm p}}$ and $\qphi_{\rm r} = \qu_{\widetilde{\rm H}{\rm c}}$, referred to as optimal beamforming. For UE-CoMIMO, we refer to Structure 1 with $\qphi_{\rm s1} = \qone$. To ensure a fair comparison between Structures 1 and 2, we normalize the power of both structures in all cases with $\tr(\qPhi_{\rm s}) = N_{\rm c}$.

\begin{figure*}
\centering
\includegraphics[width=7.00in]{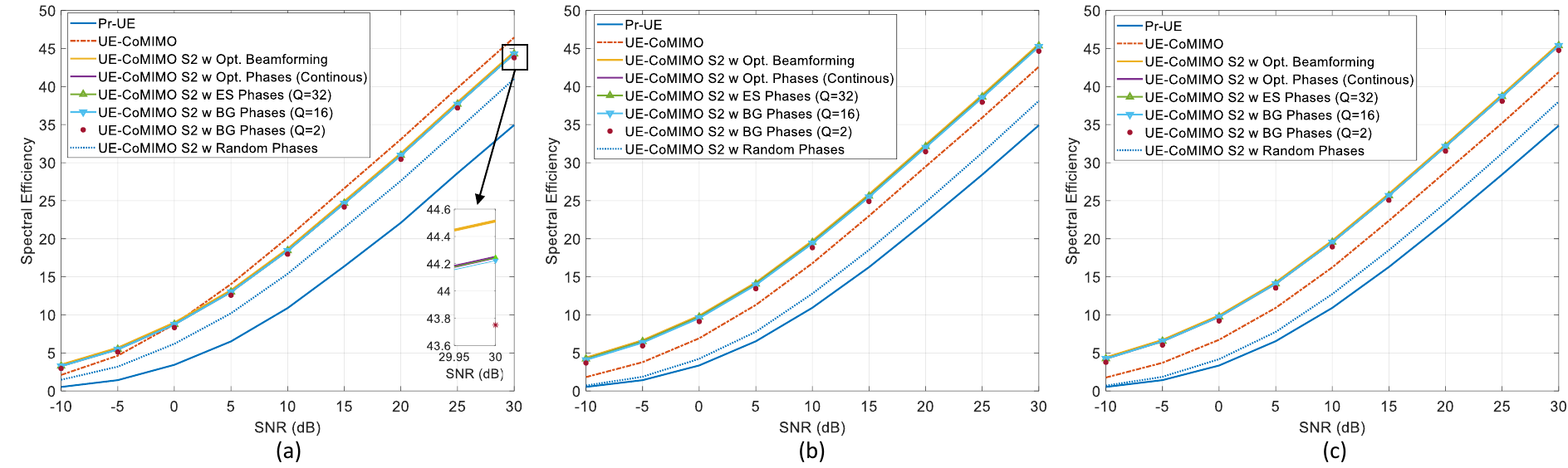}
\caption{SEs of various algorithms, particularly for Structure 2, in three different positions: (a) $(1,1,1)\,\lambda_{\rm H}$, (b) $(5,5,5)\,\lambda_{\rm H}$, and (c) $(10,10,10)\,\lambda_{\rm H}$. The algorithms include the optimal beamforming, the SDR for the optimal (continuous) phases, ES for discrete phases, and the BG algorithm for discrete phases under $Q = 32$, $16$, and $2$ levels.}
\label{fig:Results_S2}
\end{figure*}

Fig.~\ref{fig:Results_S2} shows the SEs for all the mentioned algorithms in three different Co-UE positions: (a) $(1,1,1)\,\lambda_{\rm H}$, (b) $(5,5,5)\,\lambda_{\rm H}$, and (c) $(10,10,10)\,\lambda_{\rm H}$. Both UE-CoMIMO structures significantly outperform the Pr-UE system alone. In Structure 2, algorithms employing either optimal beamforming or optimal continuous/discrete phases produce comparable results. Random phase settings perform poorly, while the BG algorithm proves highly effective, closely approximating optimal beamforming performance---even with only two phase levels ($Q = 2$). Moreover, the BG algorithm does not require full CSI, making it practical and computationally efficient.
When the Co-UE is positioned at $(1,1,1)\,\lambda_{\rm H}$, $\qH_{\rm p}$ is full rank. Fig.~\ref{fig:Results_S2}(a) shows that Structure 1 in UE-CoMIMO outperforms Structure 2, even without phase optimization. However, as shown in Figs.~\ref{fig:Results_S2}(b) and \ref{fig:Results_S2}(c), Structure 2 surpasses Structure 1, particularly as $\qH_{\rm p}$ approaches a rank-1 channel.

Notably, solving the optimization problems in (P2-1) and (P2-2) using SDR relaxes the rank-one constraint. Our simulations show that 100\% of $\Omega_{\rm t}$ and 99.8\% of $\Omega_{\rm r}$ are rank-one, highlighting the efficiency of SDR for these optimization problems. Cases where $\Omega_{\rm r}$ deviates from rank-one generally occur when the Co-UE is positioned at $(1,1,1)\,\lambda_{\rm H}$ and under high SNR conditions. This result is expected because aligning to the principal eigenvector of $\widetilde{\qH}_{\rm c}$ is often the optimal direction for (P2-2) at low SNRs, while diversity becomes more advantageous at high SNRs.

\begin{figure*}
\centering
\includegraphics[width=7.00in]{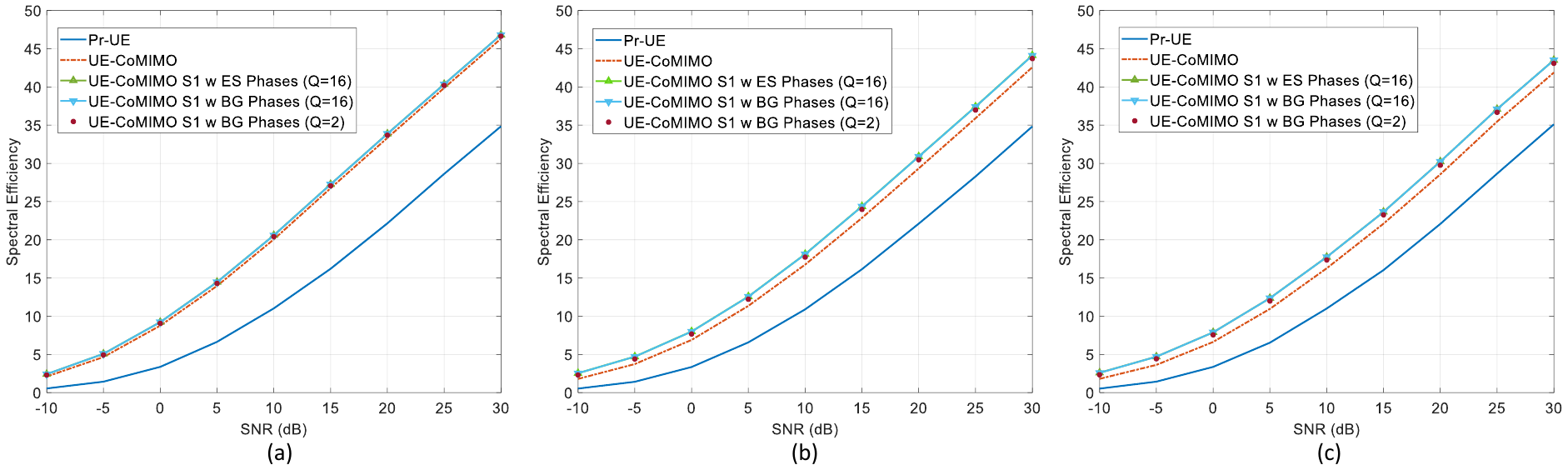}
\caption{SEs of various algorithms, particularly for Structure 1, in three different positions: (a) $(1,1,1)\,\lambda_{\rm H}$, (b) $(5,5,5)\,\lambda_{\rm H}$, and (c) $(10,10,10)\,\lambda_{\rm H}$. The algorithms include ES for discrete phases, and the BG algorithm for discrete phases under $Q = 16$ and $2$ levels.}
\label{fig:Results_S1}
\end{figure*}

We now shift our focus to Structure 1. Fig.~\ref{fig:Results_S1} shows the SEs through the ES and BG algorithms. Adjusting phases in Structure 1 yields only minor gains, particularly when $\qH_{\rm p}$ exhibits isotropic characteristics. Conversely, phase adjustments in Structure 1 become more beneficial as $\qH_{\rm p}$ approaches a rank-1 channel. This observation aligns with the analysis following Theorem \ref{theorem:special_case_S2}.\footnote{Our simulations reveal that the optimization solution of $\qOmega_{\rm s1}$ in (P1) generally deviates from rank-one when
$\qH_{\rm p}$ is isotropic. When $\qH_{\rm p}$ approaches a rank-1 channel, the optimization solution of $\qOmega_{\rm s1}$ tends to be rank-one at low SNRs. This behavior further explains the limited gains observed for Structure 1.} Despite these insights, a comparison of Figs.~\ref{fig:Results_S2}(c) and \ref{fig:Results_S1}(c) shows that Structure 2 with optimal phases outperforms Structure 1 with optimal phases. On the other hand, comparing Figs.~\ref{fig:Results_S2}(a) and \ref{fig:Results_S1}(a) reveals that Structure 1 without phase adjustments already surpasses Structure 2 with optimal phases.

Finally, we evaluate these characteristics across various $\kappa$-factors. While detailed results are omitted here for brevity, the findings align with our previously discussed conclusions.
{\bf Consequently, we find that the choice between Structures 1 and 2 depends on the specific use case. For scenarios where $\qH_{\rm p}$ is close to full rank, Structure 1 is preferred. In contrast, for scenarios where $\qH_{\rm p}$ approaches rank-1, Structure 2 is more suitable. Phase optimization offers minor gains for Structure 1 but becomes crucial for Structure 2.}

\subsection{Evaluation Through Ray-Tracing in Indoor Scenario}
\label{sec:Ray-Tracing-Indoor}

\begin{figure}
\centering
\includegraphics[width=3.35in]{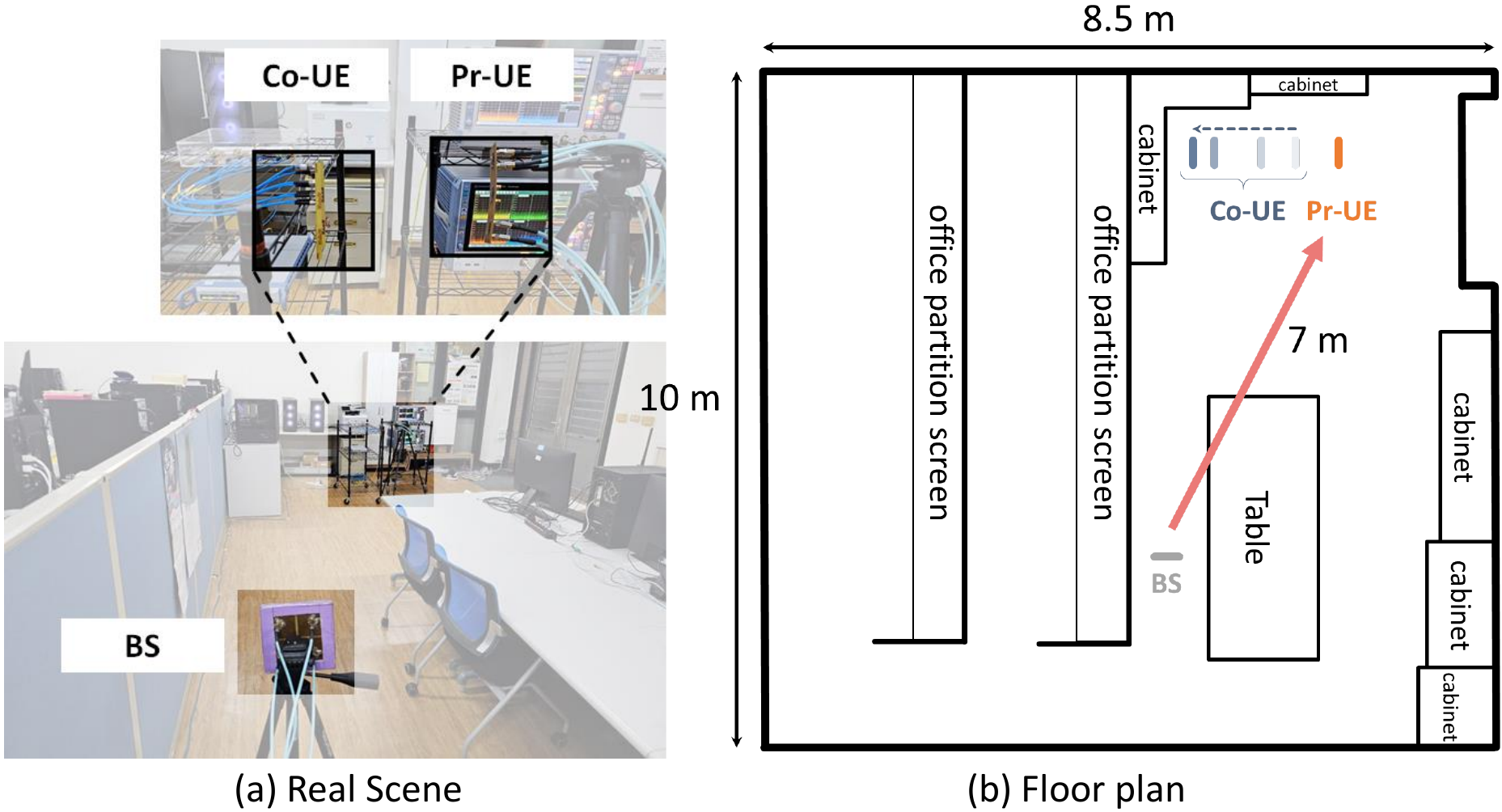}
\caption{The testing scenario: (a) real scene; (b) floor plan}
\label{fig:TestingScenario}
\end{figure}

\begin{table}[t]
\centering
\caption{Experiment parameters.} \label{tab:Setup_Exp}
\begin{tabular}{|l|l|l|}
\hline
\textbf{Environment}                    & Indoor Office  & Boston   \\ \hline
\multirow{2}{*}{\textbf{Carrier Freq.}} & $f_{\rm L} = 3.65$ GHz & $f_{\rm L} = 3.5$ GHz \\
                                        & $f_{\rm H} = 7.65$ GHz & $f_{\rm L} = 7$ GHz \\ \hline
\textbf{Bandwidth}                      & \multicolumn{2}{l|}{100 MHz}                  \\ \hline
\textbf{FFT, Subcarriers}               & \multicolumn{2}{l|}{2048/1620}                 \\ \hline
\textbf{Sampling Rate}                  & \multicolumn{2}{l|}{122.88 Msps}                   \\ \hline
\multirow{3}{*}{\textbf{Antenna Conf.}} & $M= 4$ & $M= 8$ \\
                                        & $N_{\rm c} = 4$ & $N_{\rm c} = 4$ \\
                                        & $N_1=N_2=2$   & $N_1=N_2=2$       \\ \hline
\textbf{BS and Pr-UE Distance}  & $\sim 7$ m  & $1.3-587.6$ m              \\ \hline
\multirow{2}{*}{\textbf{Pr-UE and Co-UE Distance}}  & 0.03, 0.05, & \multirow{2}{*}{0.05 m} \\
                                        & 0.1, 0.2, $\ldots$ , 0.5 m &  \\ \hline
\textbf{Tx Power}  & -12 dBm  & 0 dBm \\ \hline
\textbf{Noise Power}  & \multicolumn{2}{l|}{-94 dBm} \\ \hline
\textbf{LNA Gain}  & \multicolumn{2}{l|}{20 dB} \\ \hline
\textbf{BS Height}  &  2 m & 15 m \\ \hline
\textbf{UE Height}  &  \multicolumn{2}{l|}{1.2 m} \\ \hline
\textbf{PS Opt. Alg.}  & \multicolumn{2}{l|}{BG (4 levels)}    \\ \hline
\end{tabular}
\end{table}

\begin{figure}
\centering
\includegraphics[width=3.35in]{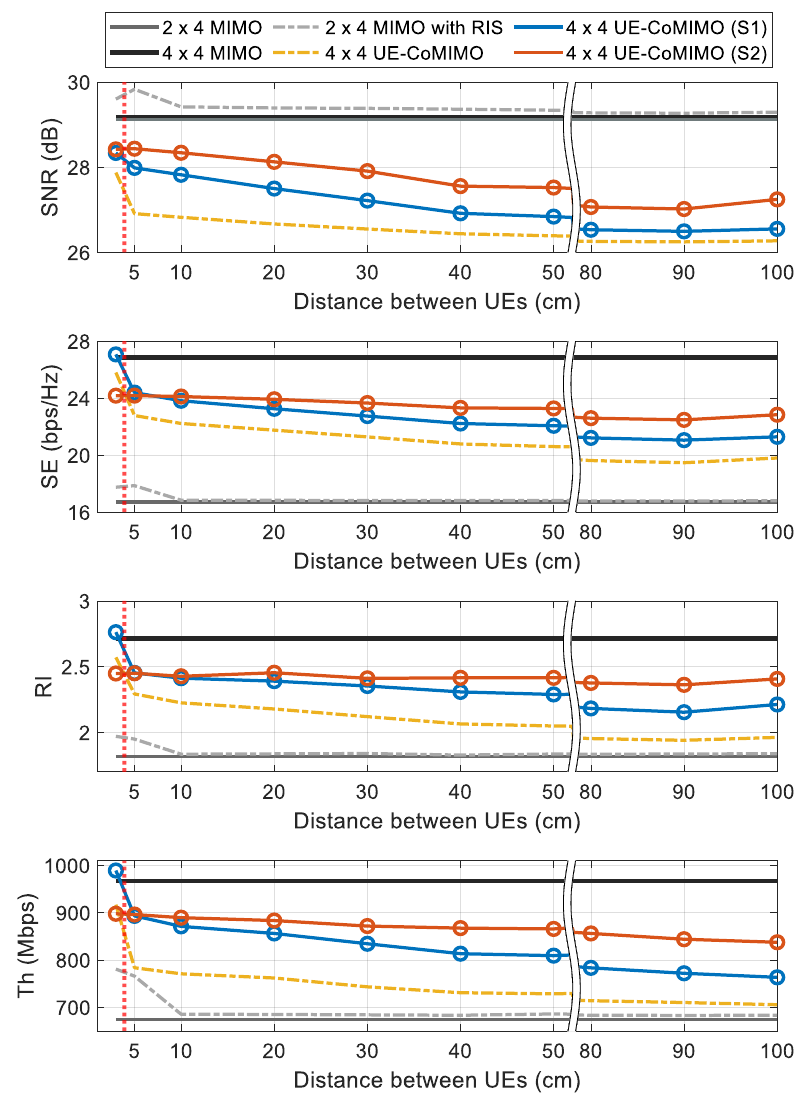}
\vspace{-0.2cm}
\caption{Performance with varying distances between Co-UE and Pr-UE. The SNRs for $2\times 4$ MIMO and $4\times 4$ MIMO are close and difficult to distinguish in the figure.}
\label{fig:Results_RT_Indoors}
\end{figure}

To gain practical insights, we conduct ray-tracing simulations. Unlike previously used statistical channel models, the ray-tracing method more accurately captures the geometric characteristics of the channels, including path loss and scattering effects. We first considered an indoor laboratory scenario in the NSYSU Electrical and Computer (EC) Engineering Building, as depicted in Fig.~\ref{fig:TestingScenario}. Section \ref{sec:Experiments} will discuss over-the-air (OTA) tests conducted in the same setting. The room dimensions are 10 m in length, 8.5 m in width, and 2.2 m in height. The BS and Pr-UE are positioned as illustrated in Fig.~\ref{fig:TestingScenario}, while the Co-UE is placed at various distances from the Pr-UE. A 3 cm distance mimics a scenario where a portable charger serves as the Co-UE, whereas 50--100 cm distances simulate use cases involving wearable devices or customer-premises equipment.

The frequency bands analyzed are $f_{\rm L}$ at 3.65 GHz, representative of current 5G NR commercial networks, and $f_{\rm H}$ at 7.65 GHz. We use Wireless InSite$^\circledR$ to model the propagation channels between the BS and Pr-UE, BS and Co-UE, and Co-UE and Pr-UE. The system uses a 5G NR MIMO-OFDM waveform with 100 MHz bandwidth and 60 kHz subcarrier spacing, consistent with the 5G NR frame structure. The BS has four antennas, while Pr-UE has four antennas: two for $f_{\rm L}$ and two for $f_{\rm H}$. The Co-UE has four antennas for both frequency bands. Table \ref{tab:Setup_Exp} summarizes the system parameters. As a benchmark, we consider the 4R4T MIMO (Pr-UE only) system, where Pr-UE operates with four antennas for $f_{\rm L}$ without UE-CoMIMO. Note that achieving this benchmark MIMO system is not feasible for a compact Pr-UE due to constrained antenna spacing. Additionally, we evaluate an active RIS-type Co-UE, which is identical to Structure 1 but lacks a frequency translation mixer. We refer to this setup as MIMO with RIS. The LNA gain settings for active RIS and Co-UE remain identical to ensure a fair comparison.

Fig.~\ref{fig:Results_RT_Indoors} shows performance trends with varying distances between Co-UE and Pr-UE. We assess SNR (in dB), spectral efficiency (SE in bps/Hz), rank indicator (RI), and throughput (TP in Mbps) for 2R4T MIMO (Pr-UE only), 4R4T MIMO (Pr-UE only), 2R4T MIMO with RIS, and 4R4T UE-CoMIMO configurations. SNR values are averaged across all receiving antennas. SE is calculated as defined in \eqref{eq:Cap1}, and TP is evaluated per 3GPP TR 38.306 \cite{3GPP-TR-38.306}, with BS-applied precoding for CQI and RI as per 5G NR standards.

The results in Fig.~\ref{fig:Results_RT_Indoors} show that the benchmark 4R4T MIMO (solid black line) achieves the highest performance, while 2R4T MIMO (solid gray line) represents the lower bound. With RIS assistance, 2R4T MIMO (gray dash-dotted line) shows modest improvements across all metrics, though these gains diminish as the distance between the RIS and Pr-UE increases. This is expected, as the four-element RIS provides limited contribution, and its impact diminishes rapidly due to propagation path loss.

Comparing 2R4T MIMO Pr-UE only (solid gray line) with 4T4R UE-CoMIMO (yellow dash-dotted line), UE-CoMIMO generally achieves better results for all metrics except SNR. SNR decreases as the distance between Co-UE and Pr-UE increases due to path loss. In the Pr-UE-only configuration, two Rx antennas operate at $f_{\rm L}$, while in the UE-CoMIMO configuration, two Rx antennas operate at $f_{\rm L}$ and two at $f_{\rm H}$. The SNR at $f_{\rm L}$ remains unaffected by the Co-UE and Pr-UE distance, so variations in UE-CoMIMO SNR are attributed to changes at $f_{\rm H}$. The decline in average SNR for UE-CoMIMO reflects lower SNRs at $f_{\rm H}$, which is expected since free-space path loss at 7.65 GHz exceeds 30 dB at 0.1 m, and the 20 dB LNA gain does not fully offset this loss. Despite lower SNRs at $f_{\rm H}$, UE-CoMIMO achieves higher SE compared to Pr-UE only. The RI for Pr-UE only remains below two, whereas UE-CoMIMO exceeds a rank of two due to additional channels provided by Co-UE. As the distance between Co-UE and Pr-UE increases, the SE, RI, and TP of UE-CoMIMO converge to those of the Pr-UE-only system.

When Co-UE and Pr-UE are in close proximity, UE-CoMIMO yields significant gains in SE, RI, and TP. At short distances, the channel matrix $\qH_{\rm p}$ approaches full rank, resulting in marked performance enhancements. However, when the distance between Co-UE and Pr-UE surpasses the Fraunhofer distance---indicated by the red dashed vertical line in Fig.~\ref{fig:Results_RT_Indoors}---the channel $\qH_{\rm p}$ degrades to a rank-one channel, reducing performance gains. {\bf In real OTA tests, described in Section \ref{sec:Experiments}, this distance is longer,  likely due to the presence of more scatterers compared to ray-tracing simulations, where path reflections and edge diffractions are simplified to reduce simulation time.}

Phase optimization using the BG algorithm improves performance for both Structures 1 and 2. Structure 2 shows significant SNR enhancements over Structure 1. At a distance of 3 cm, SE, RI, and TP for Structure 2 are initially lower than those for UE-CoMIMO due to $\qH_2$ being reduced to a rank-one channel. However, as the distance increases, the benefits of Structure 2 become more significant, aligning with previous observations. Notably, at 3 cm, Structure 1 even slightly outperforms the benchmark 4R4T MIMO, likely due to reduced channel correlations caused by spatial separation between Co-UE and Pr-UE antennas compared to the co-located antennas in the 4R4T MIMO benchmark.

Finally, although RIS also employs the BG algorithm for phase optimization, it lacks the capability to establish additional RF links in $f_{\rm H}$ that are distinguishable from the links between the BS and the Pr-UE in $f_{\rm L}$. This limitation restricts its ability to upgrade the 2R4T MIMO system to a 4R4T configuration. As a result, the performance of the 2R4T MIMO with RIS rapidly converges to that of the baseline system without RIS when the RIS-Pr-UE distance exceeds 10 cm.

\subsection{Evaluation Through Ray-Tracing in Outdoor Scenario}
\label{sec:Ray-Tracing-Outdoor}

\begin{figure*}
\centering
\includegraphics[width=7.50in]{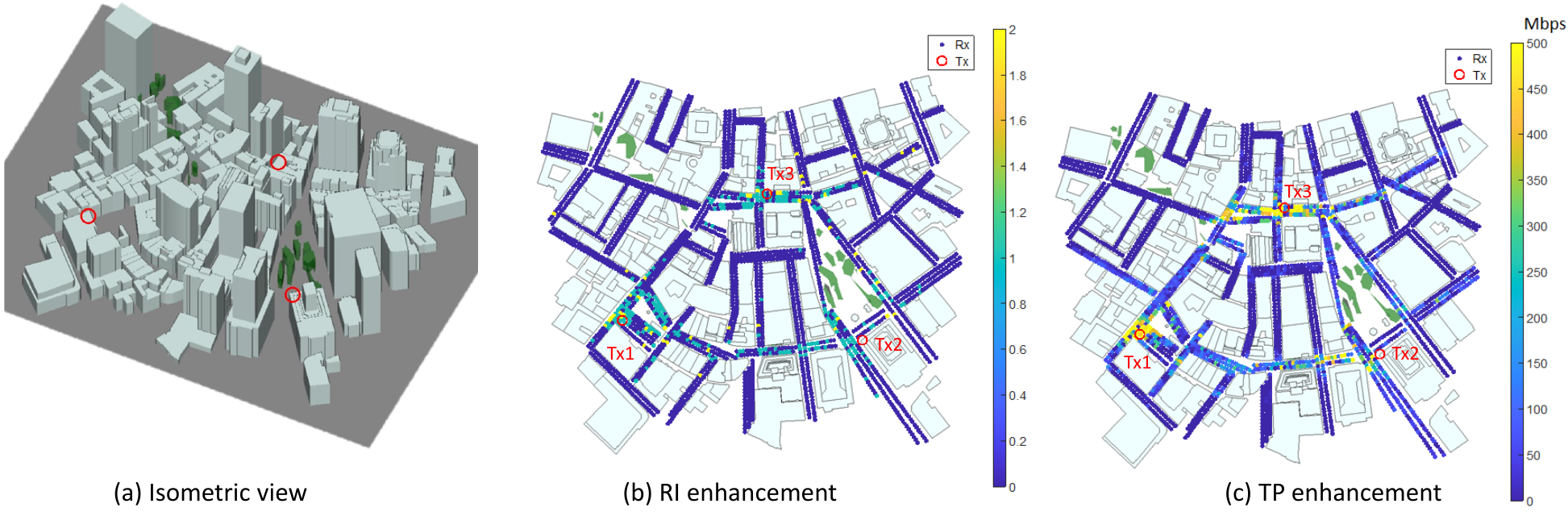} \vspace{-0.2cm}
\caption{(a) Isometric view of a section of downtown Boston in Wireless InSite \cite{WirelessInSite-Boston}. (b) Enhancement in terms of RI. (c) Enhancement in terms of TP.}
\label{fig:RT-RI-TP-HotMaps}
\end{figure*}

\begin{figure}
\centering
\includegraphics[width=3.50in]{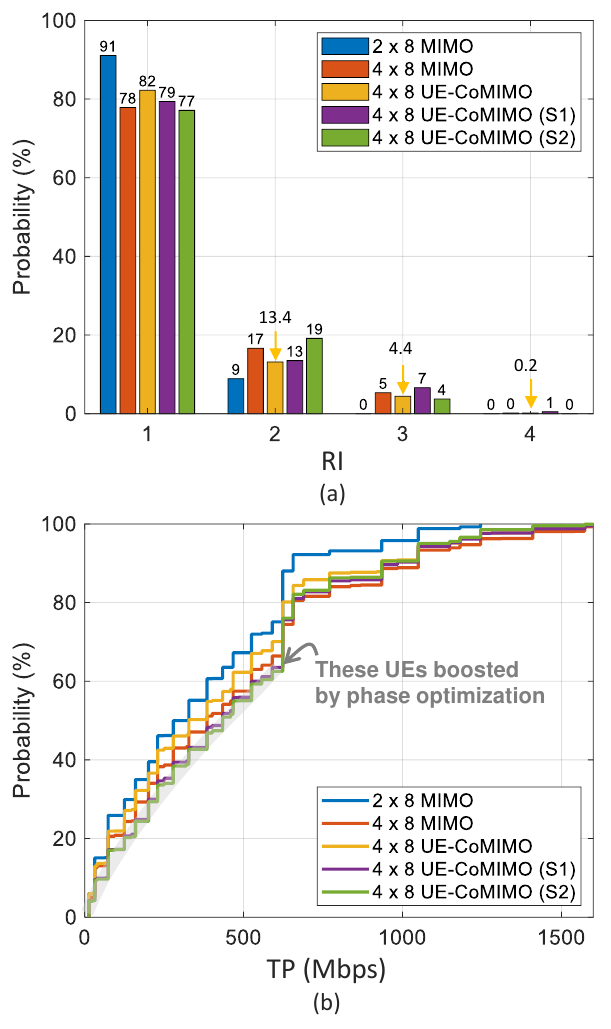} \vspace{-0.2cm}
\caption{(a) Probability distribution of the RI used. (b) CDF of the TP.}
\label{fig:RT-RI-TP-CDF}
\end{figure}

To broaden the evaluation, we conduct ray-tracing simulations for UE-CoMIMO in an outdoor area based on a section of downtown Boston. The building and foliage geometry are imported from a high-resolution shapefile developed and released by Remcom \cite{WirelessInSite-Boston}, as shown in Fig.\ref{fig:RT-RI-TP-HotMaps}(a). Each Tx represents a BS (marked as ``$\circ$'' in Fig.~\ref{fig:RT-RI-TP-HotMaps}(a)) and uses a $1 \times 8$ uniform linear array with half-wavelength spacing between antenna elements, where each antenna element is isotropic. The antenna configurations for the Co-UE and Pr-UE remain the same as those described in the previous subsection. Other system parameters are summarized in Table \ref{tab:Setup_Exp}. The UE is uniformly placed on an outdoor x-y grid with 5 m spacing between adjacent points, resulting in 2,454 UEs in the area. The height of all user grids is set to 1.2 m. There are three BSs operating on different frequency channels around 3.5 GHz, and each UE connects to the BS with the strongest SNR.

We evaluate the TP in Mbps for the 2R8T MIMO (Pr-UE only) and 4R8T UE-CoMIMO systems, with precoding applied by the BS to determine the CQI and RI as per the 5G NR standard. The rank number and TP of the 4R8T UE-CoMIMO system are higher than those of the 2R8T MIMO system. Figs.~\ref{fig:RT-RI-TP-HotMaps}(b) and \ref{fig:RT-RI-TP-HotMaps}(c) show enhancements in RI and TP, respectively. The 4R8T UE-CoMIMO system significantly boosts the rank number, particularly for UEs close to the BS, typically adding one additional rank across most areas and two additional ranks in certain regions. However, the rank enhancement diminishes when the UE is far from the BS. Additionally, in areas near Tx2, such as an open garden with limited propagation paths, rank enhancement is minimal. Despite these limitations, UE-CoMIMO still significantly improves TP, indicating that even when spatial multiplexing gains are limited, it enhances performance through mechanisms such as improved signal quality and diversity.

Fig.~\ref{fig:RT-RI-TP-CDF}(a) shows the probability distribution of the RI based on ray-tracing simulations. In the 2R8T MIMO system, 91\% of UEs have a rank of 1, and only 9\% support up to rank 2. Using the 4R8T UE-CoMIMO, the proportion of rank 1 UEs decreases to 82\%. With Structure 1 and Structure 2, this proportion further drops to 79\% and 77\%, respectively, indicating a transition to higher ranks. Structure 2 supports up to 3 ranks, while Structure 1 and UE-CoMIMO can support up to rank 4. Structure 1 exhibits a higher probability of achieving rank 4, demonstrating that phase optimization further enhances spatial multiplexing gains.

Fig.~\ref{fig:RT-RI-TP-CDF}(b) presents the cumulative distribution function (CDF) of the TP. The CDFs of TP for Structures 1 and 2 are similar and outperform those of UE-CoMIMO and 2R8T MIMO. Compared to UE-CoMIMO without phase optimization, over 60\% of UEs experience substantial performance improvements due to phase optimization. Moreover, UE-CoMIMO with phase optimization enables these UEs to achieve better TP than the benchmark 4R8T MIMO (Pr-UE only) system. This improvement is attributed to the spatial separation between the Co-UE and Pr-UE antennas, which reduces channel correlations compared to the benchmark 4R8T MIMO. This finding aligns with the observations made in Fig.~\ref{fig:Results_RT_Indoors}.

\subsection{Active RIS, UE-CoMIMO, and MIMO}
\label{sec:Comps}

In the previous subsections, we provide detailed comparisons of the performance differences among active RIS, UE-CoMIMO, and benchmark MIMO in both indoor and outdoor scenarios. Table~\ref{tab:RIS_UE-CoMIMO_MIMO} summarizes these comparisons from a top-level perspective.

First, we clarify the distinction between the active RIS used in this paper and the network-controlled repeaters (NCRs) defined in 3GPP Release 18 \cite{Wen-24COMSTMAG,Silva-24ArXiv}. NCRs are standard amplify-and-forward repeaters with beamforming capabilities, capable of receiving and processing control information from the network. The primary similarity between active RISs and NCRs is that both nodes forward signals immediately in desired directions without decoding them. However, unlike NCRs, active RISs amplify signals by only a few dB, whereas NCRs can provide amplification gains of up to 90-100 dB \cite{Astrom-24ArXiv}. This difference allows RISs to introduce minimal noise and latency, offering an advantage over NCRs. In Section \ref{sec:Ray-Tracing-Indoor}, we evaluate an active RIS-type Co-UE resembling Structure 1 but without a frequency translation mixer. While the active RIS considered in this paper enhances the SNR for the Pr-UE, it provides limited multiplexing improvement to the Pr-UE. Moreover, these gains diminish quickly as the distance between the RIS and Pr-UE increases.

Although both RIS and Co-UE assist the Pr-UE via nearby external devices, UE-CoMIMO leverages Co-UE to create additional RF links, enhancing the multiplexing capability of the Pr-UE. As demonstrated in Fig.~\ref{fig:Results_RT_Indoors}, the average SNR for UE-CoMIMO often decreases due to higher path loss or interference at higher frequency bands. Nevertheless, the multiplexing capability of UE-CoMIMO improves significantly. In contrast, the benchmark MIMO system achieves similar multiplexing gains by directly adding extra Rx antennas to the Pr-UE, maintaining consistent SNR across all Rx antennas without relying on external devices. Benchmark MIMO technology is straightforward and mature. However, this approach may not always be feasible due to practical constraints, particularly those related to the device's form factor.

\begin{table*} 
\centering
\caption{Comparison of Active RIS, UE-CoMIMO, and MIMO}
\label{tab:RIS_UE-CoMIMO_MIMO}
\begin{tabular}{llll}
\toprule
\textbf{Aspect}      & \textbf{Active RIS}                          & \textbf{UE-CoMIMO}                             & \textbf{MIMO (Extra Rx Antennas)}                                   \\ \midrule
\multirow{2}{*}{\textbf{Operation}}      & External device, & External device, amplify-and-forward      & Direct signal processing at Rx                 \\
                                         & amplify-and-forward & with frequency translation       &                  \\ \midrule
\textbf{Latency/Noise Introduction}      & Low                                      & Low (may increase with extra mixer)             & Minimal                        \\ \midrule
\textbf{Multiplexing Capability}         & Limited       & High (via Co-UE)               & High (via additional Rx antennas)             \\ \midrule
\multirow{2}{*}{\textbf{Advantages}}     & Enhances SNR,             & Higher SE and diversity      & High SE, mature technology \\
                                         & no change in Rx design    &                              &   \\ \midrule
\multirow{3}{*}{\textbf{Disadvantages}}  & Limited range, & Distance-sensitive, & Limited applicability \\
                                         & minimal multiplexing improvement, & requires Co-UE coordination    & for compact devices  \\
                                         & coordination required &                       &   \\ \bottomrule
\end{tabular}
\end{table*}

\section{Proof of Concept} \label{sec:Experiments}

\begin{figure*}
\centering
\includegraphics[width=7.50in]{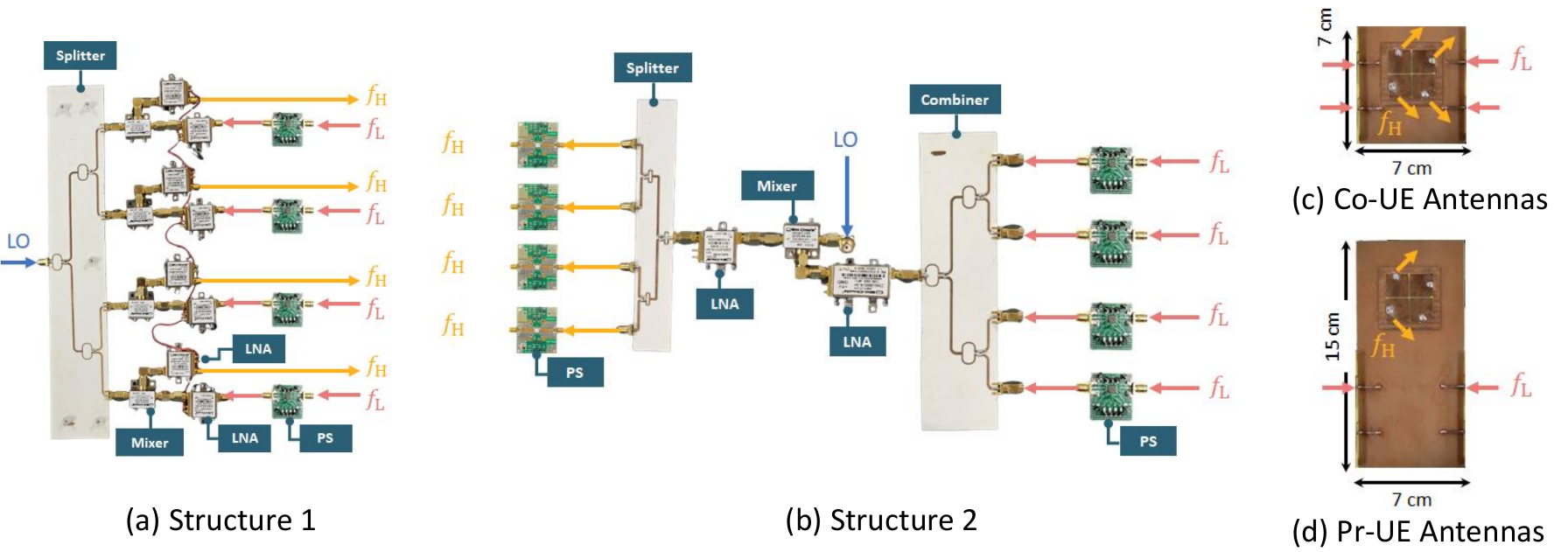} 
\caption{Hardware configurations of (a) Structure 1 and (b) Structure 2, following the architectures depicted in Fig.~\ref{fig:BF_structure}. Antenna configurations for (c) Co-UE and (d) Pr-UE are shown.}
\label{fig:L1_Relay_RealPhoto}
\end{figure*}

In this section, we evaluate the feasibility and performance of the UE-CoMIMO system through OTA testing. The test environment, configuration, and frequency bands for $f_{\rm L}$ and $f_{\rm H}$ are consistent with those used in the ray-tracing simulations for the indoor scenario discussed in Section \ref{sec:Ray-Tracing-Indoor}. We employ a 5G NR MIMO-OFDM system, utilizing the parameters listed in Table \ref{tab:Setup_Exp} (Indoor Office). The system adheres to 5G NR standards, ensuring that the experimental results are representative of real-world 5G communications.

\subsection{Device and Implementation} \label{sec:Device&Implementation}

Following the architecture shown in Fig.~\ref{fig:BF_structure}, we implemented both Structure 1 and Structure 2, as shown in Fig.\ref{fig:L1_Relay_RealPhoto}. Component specifications are detailed in Table \ref{tab:Power_RFElement}. The Co-UE is equipped with four antennas along the side edge of the phone for the $f_{\rm L}$ band and another four antennas on the back cover for the $f_{\rm H}$ band, as shown in Fig.~\ref{fig:L1_Relay_RealPhoto}(c). These antennas are designed to fit within the form factor of a 5G smartphone, taking space constraints into account. The side-edge antennas at the low mid-band are nearly omnidirectional, while the back-cover antennas at the upper mid-band are slightly directional. For phase control, we use the {\tt MAP-010144} digital phase shifter (DPS) for the $f_{\rm L}$ band and the {\tt TGP2105-SM} DPS for the $f_{\rm H}$ band. Both are 6-bit DPSs capable of phase shifts ranging from 0 to 360 degrees in 5.625-degree increments. However, as discussed in Section \ref{sec: sim_adv_UECoMIMO}, performance with $Q=2$ is already close to that of continuous phase shifts. To balance performance and complexity, we adopt four discrete phase levels ($Q=4$), i.e., $0$, $90$, $180$, and $270$ degrees.

For testing, we use a transmitter and receiver to serve as the BS and Pr-UE, respectively. The BS employs four one-port Rohde \& Schwarz (R\&S) vector signal generators ({\tt SGT100A}) connected to four transmit antennas. Multiple spatial-stream 5G OFDM signals are transmitted via precoding in the 3.6--3.7 GHz band over the air. At the Pr-UE, signals are received at $f_{\rm L}$ and $f_{\rm H}$ using two mobile 5G antennas and two upper mid-band antennas, as shown in Fig.~\ref{fig:L1_Relay_RealPhoto}(d). These received signals from the four antennas (two at $f_{\rm L}$ and two at $f_{\rm H}$) are downconverted to baseband using a four-port R\&S digital oscilloscope ({\tt RTP164}).

Baseband signal processing is performed on a personal computer using a C program. Synchronization is achieved through 5G synchronization signal blocks (SSB). Specifically, the BS transmits the SSB at $f_{\rm L}$, while the Pr-UE receives the SSB at both $f_{\rm L}$ and $f_{\rm H}$. The $f_{\rm H}$ band signals are frequency-translated by the Co-UE. Synchronization is performed independently at $f_{\rm L}$ and $f_{\rm H}$. OTA tests reveal that the estimated time differences between the two frequencies are on the order of a few nanoseconds, indicating that the link through the Co-UE introduces negligible latency. In addition to the SSB, other reference signals, such as downlink demodulation reference signals (DMRS) and data payload, are transmitted from the BS and frequency-translated by the Co-UE. The estimated channels based on DMRS from both frequencies are utilized to calculate average SNR (dB), SE (bps/Hz), RI, and TP (bps) for the 4T4R MIMO and 4T8R UE-CoMIMO systems. These metrics align with those used in ray-tracing simulations.

The Pr-UE computes SE using \eqref{eq:Cap1} based on the estimated channels from DMRS. The PSs at the Co-UE are controlled by the Pr-UE through the BG algorithm. Communication between the Pr-UE and Co-UE occurs over a half-duplex Bluetooth link. As calculated in \eqref{eq:power_consumption}, the power consumption for the Co-UE is 2420.6 mW for Structure 1 and 602.2 mW for Structure 2, both of which are suitable for terminal devices. Additionally, the Pr-UE can effectively control the on/off state of the Co-UE's forwarding functionality to further reduce the Co-UE's power consumption.

\subsection{Results}
\begin{table*}[h]
\centering
\caption{Experiment results of Pr-UE only and UE-CoMIMO for Structure 1.} \label{tab:S1_SNR_SE}
\begin{tabular}{rcccccccccccc}
\toprule
\multicolumn{1}{c}{} & \multicolumn{4}{c}{\textbf{Pr-UE Only}} & \multicolumn{4}{c}{\textbf{Structure 1 (Initial PS)}} & \multicolumn{4}{c}{\textbf{Structure 1 (BG  PS)}} \\
\cmidrule(rl){2-5} \cmidrule(rl){6-9} \cmidrule(rl){10-13}
Co-UE Dist. (m) & SNR (dB) & SE (bps/Hz) & RI (\#) & TP (Mbps) & SNR & SE & RI & TP & SNR & SE & RI & TP \\
\cmidrule(rl){1-1} \cmidrule(rl){2-5} \cmidrule(rl){6-9} \cmidrule(rl){10-13}
0.03 & \multirow{3}{*}{27.73} & \multirow{3}{*}{15.12} & \multirow{3}{*}{1.39} & \multirow{3}{*}{503.4} & 29.06 & 30.03 & 3.00 & 1275.2 & 30.33 & 30.77 & 3.00 & 1286.9 \\
0.5  &                        &                        &                       &                         & 27.12 & 28.39 & 2.94 & 1024.2 & 26.25 & 28.89 & 3.00 & 1044.6 \\
1    &                        &                        &                       &                         & 24.77 & 24.91 & 2.03 & 870.21 & 24.10 & 26.10 & 3.00 & 1151.8 \\
\bottomrule
\end{tabular}
\end{table*}

\begin{table*}[h]
\centering
\caption{Experiment results of Pr-UE only and UE-CoMIMO for Structure 2.} \label{tab:S2_SNR_SE}
\begin{tabular}{rcccccccccccc}
\toprule
 \multicolumn{1}{c}{} & \multicolumn{2}{c}{\textbf{SNR of Rx1 at $f_{\rm H}$}} & \multicolumn{2}{c}{\textbf{SNR of Rx2 at $f_{\rm H}$}} & \multicolumn{4}{c}{\textbf{Structure 2 (Initial  PS)}} & \multicolumn{4}{c}{\textbf{Structure 2 (BG  PS)}} \\
 \cmidrule(rl){2-3} \cmidrule(rl){4-5} \cmidrule(rl){6-9} \cmidrule(rl){10-13}
Co-UE Dist. (m) & Initial & BG & Initial & BG & SNR & SE & RI & TP & SNR & SE & RI & TP \\
\cmidrule(rl){1-1}  \cmidrule(rl){2-3} \cmidrule(rl){4-5} \cmidrule(rl){6-9} \cmidrule(rl){10-13}
0.03 & 26.18 & 33.59 & 22.37 & 30.50 & 23.22 & 20.78 & 2.25 & 755.2 & 29.63 & 23.34 & 2.97 & 835.7 \\
0.5  & 6.66  & 18.53 & 3.95  & 13.17 & 23.37 & 17.60 & 1.65 & 533.4 & 23.78 & 20.71 & 3.00 & 723.1 \\
1    & 3.02  & 9.88  & -1.69 & 10.56 & 23.41 & 17.31 & 1.99 & 629.9 & 23.52 & 19.37 & 2.03 & 632.4 \\
\bottomrule
\end{tabular}
\end{table*}

We conduct experimental measurements in an indoor laboratory at the NSYSU EC Engineering Building to evaluate the UE-CoMIMO system, as shown in Fig.~\ref{fig:TestingScenario}. The Co-UE is positioned at varying distances from the Pr-UE: 0.03 m, 0.5 m, and 1 m. These configurations result in a full-rank channel at 0.03 m and rank-defective channels at 0.5 m and 1 m for $\qH_{\rm p}$.

Without UE-CoMIMO, the Pr-UE supports only two spatial streams. In the initial experiment with Structure 1, the LNAs in each link of the frequency translator provide approximately 40 dB of gain. However, the free-space path loss is 50.11 dB at 1 m for $f_{\rm H} = 7.65$ GHz, meaning the LNAs cannot fully compensate for path loss at $f_{\rm H}$. The results for Structure 1 are summarized in Table \ref{tab:S1_SNR_SE}.

In most cases (except for the 0.03 m scenario), the SNRs of UE-CoMIMO decrease, indicating that the SNRs at $f_{\rm H}$ are lower than those at $f_{\rm L}$. This result is expected, as the link between the Co-UE and Pr-UE operates at a higher frequency and is therefore subject to greater path loss. Despite the lower SNRs and without applying the BG algorithm for PSs, UE-CoMIMO still provides significant enhancements in SE compared to Pr-UE, which achieves an SE of only 15.12 bps/Hz. Specifically, in the 0.03 m scenario, the SE is remarkably boosted to 30.03 bps/Hz, as the two additional channels provided by the Co-UE also enhance the rank of the system. Even at 0.5 m and 1 m, the SEs are significantly improved despite the lower average SNRs compared to Pr-UE. Comparing the RI results with those from the ray-tracing simulations in Fig.~\ref{fig:Results_RT_Indoors}, the experimental rank values are more optimistic. This discrepancy indicates that practical environments provide more scatterers, increasing the likelihood of establishing a full-rank MIMO link between the Co-UE and Pr-UE. {\bf This finding highlights the advantage of real-world scenarios where scatter-rich environments improve the performance of the UE-CoMIMO system.}

When the BG algorithm is applied, the SEs are further enhanced. In the 0.03 m scenario, both the SNR and SE are significantly improved. For the 0.5 m scenario, the SE increases while the SNR remains comparable to the case without BG. This behavior is expected because the BG algorithm for Structure 1 focuses on maximizing SE rather than SNR, resulting in an increase in RI. Similar improvements are observed in the 1 m scenario.

Next, we analyze the performance of Structure 2. Due to the lack of a gain control mechanism in the LNA for simplicity, the antenna output power of Structure 2 is lower than that of Structure 1. This discrepancy arises because Structure 2 incorporates an additional high-frequency PS. Consequently, direct comparisons of Structure 1 and Structure 2 may not be entirely fair. Therefore, we primarily focus on the performance differences within Structure 2, comparing results before and after applying the BG algorithm. Table \ref{tab:S2_SNR_SE} shows the SNRs for the two antennas at $f_{\rm H}$ for Pr-UE. After applying the BG algorithm, the SNRs for $f_{\rm H}$ are significantly improved across all three distances. This result is expected because the BG algorithm in Structure 2 primarily targets SNR optimization for the $f_{\rm H}$ channel, effectively creating four high-quality channels for the receiver. These improvements in SNR lead to corresponding enhancements in SE, RI, and TP.
Unlike Structure 1, where SE improvements from BG are modest, Structure 2 demonstrates significant SE enhancements after BG for all three distances. This observation aligns with our simulation results discussed in Section \ref{sec: sim_adv_UECoMIMO}, further validating the effectiveness of the BG algorithm in improving the performance of UE-CoMIMO systems, especially in high-frequency scenarios.

\section{Conclusion}
\label{sec:Con}

This paper analyzed the performance of the UE-CoMIMO system, which utilizes multiple fixed or portable devices within a personal area to create a virtually expanded antenna array. Our analytical results demonstrate that UE-CoMIMO significantly enhances the system's effective channel response without requiring PS optimization, with additional improvements achievable through the optimized integration of PSs into portable devices, categorized as Structures 1 and 2.
Our findings indicate that the preferred structure depends on the specific use case. For scenarios where the relay channel exhibits near full-rank characteristics, such as when a wearable device serves as the Co-UE, Structure 1 is the optimal choice. Conversely, when the relay channel approaches rank-1 characteristics, such as when customer premise equipment serves as the Co-UE, Structure 2 is more suitable. While phase optimization provides minimal benefits for Structure 1, it is critical for maximizing the performance of Structure 2.
We also implemented these collaborative devices and validated our observations in a real 5G environment. Our testing results suggest that in practical use cases, the presence of more scatterers increases the likelihood of establishing a full-rank MIMO link between the Co-UE and Pr-UE.

Future work includes extending the system to support multiple Co-UE devices, which could provide more significant gains but may also increase the algorithms' complexity. Additionally, to address interference-prone links between the Co-UE and Pr-UE, employing different code rates for additional spatial streams could prove beneficial. Expanding the UE antenna count could also enable advanced functionalities such as sensing augmentation \cite{Tsai-23COMMAG}, making the development of models for these applications an intriguing direction for future research.

\section*{Appendix A}

This appendix addresses the case from Section \ref{sec:Rank Modification} where $N_1 \geq M$. To analyze this scenario, we define the following matrices:
\begin{equation} \label{eq:def_U_H1in12}
 \qU_{{\rm H}_1} = \begin{bmatrix} \qU_{{{\rm H}_1},1} ~\qU_{{{\rm H}_1},2}  \end{bmatrix}~\mbox{and}~ \qSigma_{{\rm H}_1} = \begin{bmatrix} \qD_{{\rm H}_1} \\ \qzero  \end{bmatrix},
\end{equation}
where $\qU_{{{\rm H}_1},1} \in \bbC^{N_1 \times M}$, $\qU_{{{\rm H}_1},2} \in \bbC^{N_1 \times (N_1-M)}$, and $\qD_{{\rm H}_1} \in \bbR^{M \times M}$.
Substituting these definitions into the expression for $\qH$ in \eqref{eq:def_H,y,z}, we can rewrite $\qH$ as:
\begin{equation} \label{eq:H_inOtherForm}
 \qH = \begin{bmatrix}
        \qU_{{\rm H}_1,1} & \qzero & \qU_{{\rm H}_1,2} \\
        \qzero   & \qI_{N_2}    & \qzero
       \end{bmatrix}
       \begin{bmatrix}
       \qD_{{\rm H}_1} \\
       \qH'_2  \\
       \qzero
       \end{bmatrix}
       \qV_1^{\rm H},
\end{equation}
where $\qH'_2 = \qH_2 \qV_1 \in \bbC^{N_2 \times M}$. We then define the matrix $\qA$ as
\begin{equation} \label{eq:A_def1}
 \qA = \begin{bmatrix}
       \qD_{{\rm H}_1} \\
       \qH'_2
       \end{bmatrix} \in \bbC^{(M+N_2) \times M}.
\end{equation}
The SVD of $\qA$ can be expressed as
\begin{equation} \label{eq:A_svd1}
 \qA = \begin{bmatrix} \qU_{{\rm A},1} ~\qU_{{\rm A},2} \end{bmatrix}
       \begin{bmatrix}
       \qD_{\rm A} \\
       \qzero
       \end{bmatrix}
       \qV_{\rm A}^{\rm H},
\end{equation}
where $\qU_{{\rm A},1} \in \bbC^{(M+N_2) \times M}$, $\qU_{{\rm A},2} \in \bbC^{(M+N_2) \times N_2}$, $\qD_{\rm A} \in \bbR^{M \times M}$, and $\qV_{\rm A} \in \bbC^{M \times M}$.\footnote{
The SVD of $\qH$ is given by
\begin{equation*}
 \qH = \begin{bmatrix} \qU_{{\rm H},1} ~\qU_{{\rm H},2} \end{bmatrix}
       \begin{bmatrix}
       \qD_{\rm H} \\
       \qzero
       \end{bmatrix}
       \qV_{{\rm H}}^{\rm H},
\end{equation*}
where ${\qU_{{\rm H},1} \in \bbC^{(N_1+N_2) \times M}}$, $\qU_{{\rm H},2} \in \bbC^{(N_1+N_2) \times (N_1+N_2-M)}$, $\qD_{\rm H} \in \bbC^{M \times M}$, and $\qV_{{\rm H}} \in \bbC^{M \times M}$. The components are related as follows:
\begin{align*}
 \qU_{{\rm H},1} &=
        \begin{bmatrix}
        \qU_{{\rm H}_1,1} & \qzero   \\
        \qzero   & \qI_{N_2}
        \end{bmatrix}
        \qU_{{\rm A},1}, \\
 \qU_{{\rm H},2} &=
        \begin{bmatrix}
            \begin{pmatrix}
            \qU_{{\rm H}_1,1} & \qzero   \\
            \qzero   & \qI_{N_2}
            \end{pmatrix}
            \qU_{{\rm A},2},~
            \begin{pmatrix}
             \qU_{{\rm H}_1,2} \\
             \qzero
            \end{pmatrix}
        \end{bmatrix}, \\
 \qD_{\rm H} &= \qD_{\rm A}, ~ \qV_{{\rm H}} = \qV_1 \qV_{\rm A}.
\end{align*}}
Thus, similar to the case $N_1 < M$, computing the singular values of $\qH$ is equivalent to computing the SVD of $\qA$.
By combining \eqref{eq:A_def1} and \eqref{eq:A_svd1}, we obtain
\begin{equation} \label{eq:AA_InUpdate}
 \qA^{\rm H} \qA = \qD_{{\rm H}_1}^2 + (\qH'_2)^{ {\rm H}} \qH'_2 = \qV_{{\rm A}} \qD_{\rm A}^2 \qV_{{\rm A}}^{\rm H},
\end{equation}
which represents the rank-$N_2$ modification of $\qD_{{\rm H}_1}^2$.

{\renewcommand{\baselinestretch}{1.06}
\bibliographystyle{IEEEtran}
\bibliography{IEEEabrv,References}

\end{document}